\documentclass[aps, prx, twocolumn, a4paper, superscriptaddress, longbibliography]{revtex4-2}

\usepackage[T1]{fontenc} % Special characters. One of the cited authors in the bibliography has a \k{a} as part of their name.

\usepackage{graphicx} % Include figure files
\usepackage{dcolumn} % Align table columns on decimal point
\usepackage{amsmath, bm} % maths packages
\usepackage{hyperref} % add 
\usepackage{color} % colour in document
\usepackage{gensymb} % degree symbol
\usepackage{multibib}
\usepackage{siunitx}
\makeatletter
\g@addto@macro\bfseries{\boldmath}
\makeatother

\newcommand{\musr}{$\mu$SR}
\newcommand{\SmS}{S$\mu$S}
\newcommand{\PSI}{Paul Scherrer Institute}

\newcommand{\neel}{N{\'e}el}

\begin{document}

\title {Thermally-driven reorientation of the \neel\ vector in altermagnetic MnTe}

\author{Thomas~J.~Hicken}
\email{thomas.hicken@psi.ch}
\affiliation{PSI Center for Neutron and Muon Sciences, 5232 Villigen PSI, Switzerland}

\author{Klára~Uhlířová}
\affiliation{Faculty of Mathematics and Physics, Charles University, Prague, Czech Republic}

\author{Helena~Reichlová}
\affiliation{Institute of Physics, Czech Academy of Sciences, Prague, Czech Republic}

\author{Ondrej~Caha}
\affiliation{Department of Condensed Matter Physics, Faculty of Science, Masaryk University, Kotlářská 2, Brno, 61137, Czechia} 

\author{Vincent~C.~Morano}
\affiliation{PSI Center for Neutron and Muon Sciences, 5232 Villigen PSI, Switzerland}

\author{Oliver~Amin}
\affiliation{School of Physics and Astronomy, University of Nottingham, Nottingham, United Kingdom}

\author{Alfred~Dal~Din}
\affiliation{School of Physics and Astronomy, University of Nottingham, Nottingham, United Kingdom}

\author{J.~Hugo~Dil}
\affiliation{PSI Center for Photon Science, 5232 Villigen PSI, Switzerland}
\affiliation{Institut de Physique, École Polytechnique Fédérale de Lausanne, 1015 Lausanne, Switzerland}

\author{Marek~Duchan}
\affiliation{Department of Condensed Matter Physics, Faculty of Science, Masaryk University, Kotlářská 2, Brno, 61137, Czechia} 

\author{Dominik~Kriegner}
\affiliation{Institute of Physics, Czech Academy of Sciences, Prague, Czech Republic}

\author{Hubertus~Luetkens}
\affiliation{PSI Center for Neutron and Muon Sciences, 5232 Villigen PSI, Switzerland}

\author{Zeno~Maesen}
\affiliation{PSI Center for Neutron and Muon Sciences, 5232 Villigen PSI, Switzerland}

\author{Jan~Martinek}
\affiliation{Faculty of Mathematics and Physics, Charles University, Prague, Czech Republic}

\author{Zaher~Salman}
\affiliation{PSI Center for Neutron and Muon Sciences, 5232 Villigen PSI, Switzerland}

\author{Peter~Wadley}
\affiliation{School of Physics and Astronomy, University of Nottingham, Nottingham, United Kingdom}

\author{Oksana~Zaharko}
\affiliation{PSI Center for Neutron and Muon Sciences, 5232 Villigen PSI, Switzerland}

\author{Juraj~Krempaský}
\affiliation{PSI Center for Photon Science, 5232 Villigen PSI, Switzerland}

\author{Jonas~A.~Krieger}
\email{jonas.krieger@psi.ch}
\affiliation{PSI Center for Neutron and Muon Sciences, 5232 Villigen PSI, Switzerland}

\begin{abstract}
Altermagnets are novel magnetic systems that possess a spin-polarized electronic band structure without a net magnetic moment, making them promising for device applications.
Hexagonal MnTe, a prototypical altermagnet, arguably exhibits the most properties consistent with theoretical predictions, including an anomalous Hall effect despite a vanishing net magnetization, and altermagnet-induced electronic band splitting.
However, fundamental questions remain, including why some effects only appear significantly below the magnetic ordering temperature.
Here, we resolve this discrepancy by revealing a reorientation of the \neel\ vector in single-crystalline MnTe.
The \neel\ vector points 30$^\circ$ from the $a$-axis at low $T$, before aligning directly with the $a$-axis around $T^*\simeq260$ K.
We attribute this to single-ion anisotropy, which depends on temperature-dependent lattice parameters.
We obtained these results using muon-spin spectroscopy, magnetization measurements, and X-ray diffraction; we show that the findings are consistent with neutron diffraction.
Manipulating this effect, for example through strain, could unlock sensitive electronic detection schemes for external stimuli, paving the way for functional altermagnetic devices.
\end{abstract}

\date{\today}
\maketitle

Magnetically ordered solids are traditionally divided into two elementary phases: ferromagnets and antiferromagnets.
A key characteristic of ferromagnets is a macroscopic magnetization due to the alignment of magnetic moments, which leads to a spin-polarized electronic structure and the anomalous Hall effect (AHE).
Conversely, collinear antiferromagnets lack a net magnetization due to compensated magnetic order, yielding a spin-degenerate band structure.
Recently, altermagnetism, a third fundamental collinear magnetic phase that combines aspects of both other phases, was discovered~\cite{smejkal2022beyond}.

Altermagnetism is characterised by a compensated collinear magnetic order in real space, specifically opposite spin sublattices connected by crystal rotation symmetries (symmorphic/non-symmorphic, proper/improper).
In reciprocal space, unconventional spin-polarized bands with corresponding rotational symmetries manifest.
This real-to-reciprocal-space correspondence produces electronic bands with broken time-reversal symmetry and alternating momentum-dependent spin splitting~\cite{smejkal2022emerging}.
Despite the compensated magnetic structure, these properties give rise to new phenomena, underpinning fundamental science and enabling robust spintronic memories, charge-spin conversion devices, and optospintronics~\cite{bai2024altermagnetism}.

Hexagonal MnTe has recently attracted significant attention as a prototypical altermagnet~\cite{smejkal2022beyond,smejkal2022emerging,krempasky2024altermagnetic,kriegner2016multiple,amin2024nanoscale,yumnam2024magnon,kluczyk2024coexistence,hariki2024x,gonzalez_betancourt_anisotropic_2024,lee2025dichotomous,lee2024broken}.
Below its ordering temperature of $T_\text{N}\simeq310$~K, it consists of ferromagnetically ordered Mn layers stacked such that magnetic moments in alternating layers point in opposite directions~\cite{kunitomi1964neutron,szuszkiewicz2005neutron,kriegner2017magnetic}.
These two spin sublattices are related by a six-fold crystal rotation combined with a half-unit-cell translation.
Despite its vanishing net magnetization and collinear spin ordering, MnTe demonstrates an AHE~\cite{wasscher1965evidence,gonzalez2023spontaneous,kluczyk2024coexistence}, which cannot occur in conventional collinear antiferromagnets.
Additionally, photoemission experiments have observed the band splitting predicted to result from the altermagnetic lifting of Kramers spin degeneracy~\cite{krempasky2024altermagnetic,lee2024broken,osumi2024observation}.
Moreover, MnTe may provide a promising platform for magnetic memory applications~\cite{kriegner2016multiple}.

Given these observations, detailed studies of the magnetism using various experimental probes are essential.
To use MnTe in device applications, microscopic control of its domains is required~\cite{amin2024nanoscale}; thus, a complete picture of the realised magnetic structure(s) is crucial.
Furthermore, surprising observations, such as the disappearance of the AHE~\cite{wasscher1965evidence,gonzalez2023spontaneous} and anisotropic magnetoresistance~\cite{kriegner2016multiple} well below the magnetic ordering temperature, hitherto lack explanation.
Despite the technologically advantageous properties of MnTe, including a magnetic ordering temperature above room temperature and the ability to grow both single-crystal and thin-film samples, this knowledge gap limits research and hinders the realization of spintronic devices.

\begin{figure}
	\centering
	\includegraphics[width=0.6\linewidth]{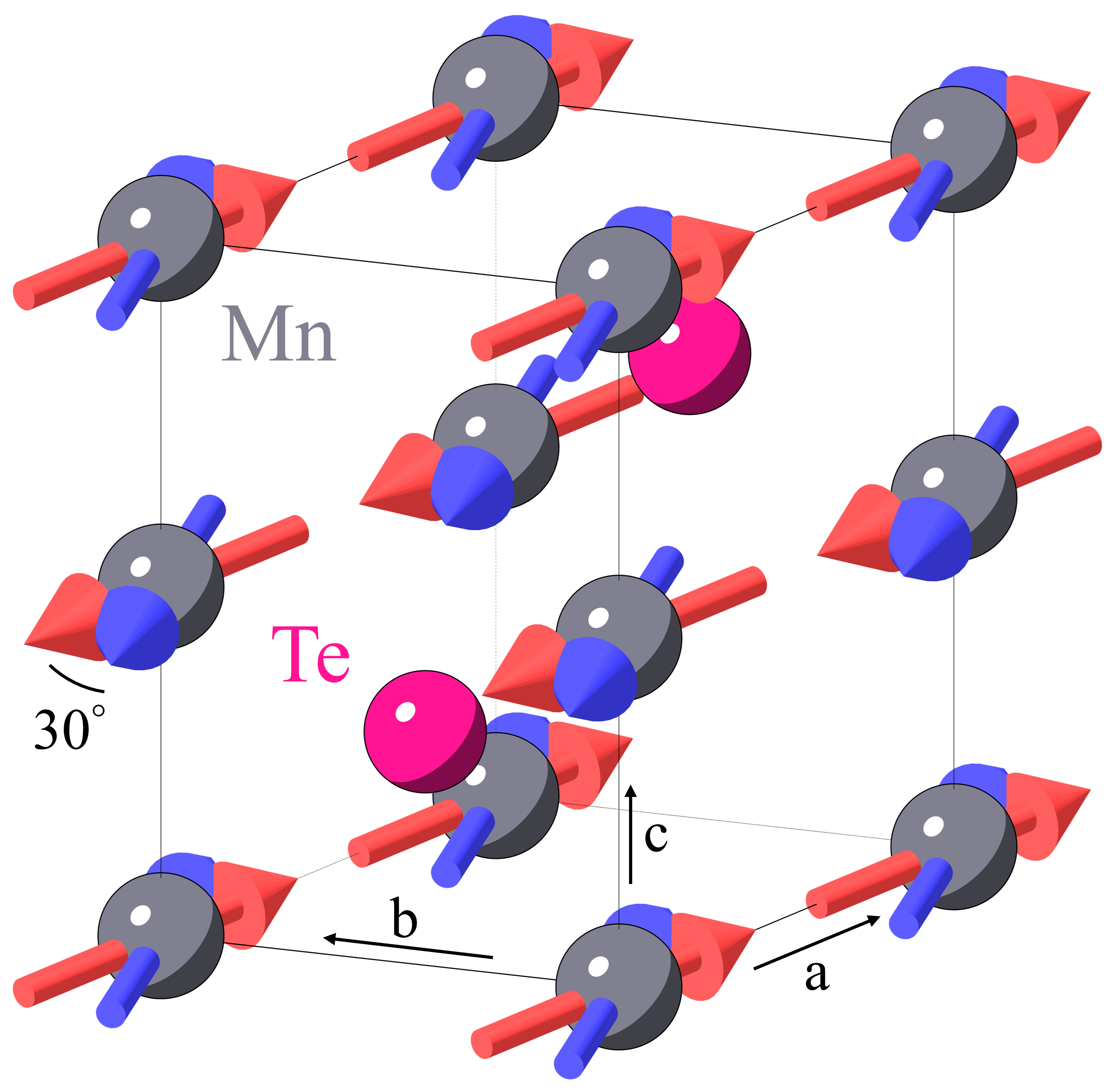}
	\caption{Magnetic structure of MnTe. The magnetic moments lie along the $a^*$-axis $\langle10\bar{1}0\rangle$ (30$^\circ$ from the $a$-axis) at low temperatures (blue arrows), before reorienting to the $a$-axis $\langle2\bar{1}\bar{1}0\rangle$ above $T^*\simeq260$~K (red arrows).}
	\label{fig:fig1}
\end{figure}

In this Letter, we employ muon-spin spectroscopy (\musr), a local probe technique highly sensitive to the precise details of the magnetic structure.
\musr\ has previously been used to study altermagnets such as Co$_{1/4}$NbSe$_2$~\cite{graham2026local}, where it confirmed a magnetic structure consistent with altermagnetism.
Here we demonstrate, through careful \musr\ measurements of single-crystalline MnTe, that the magnetic moments reorient (i.e., the \neel\ vector changes direction) around $T^*\simeq260$~K, pointing 30$^\circ$ from the $a$-axis at low $T$, and along the $a$-axis at high $T$ [Fig.~\ref{fig:fig1}].
We argue that this spin reorientation transition is a natural consequence of Mn single-ion anisotropy, with changes in the lattice parameters (likely due to thermal expansion) driving the effect.
Thus, through control of the lattice parameters (for example, by introducing strain to thin films or bulk crystals), one can control the \neel\ vector, unlocking new potential applications near or at room temperature, and novel research avenues.

Our zero-field (ZF) \musr\ measurements of single-crystal MnTe reveal three distinct temperature regions below $T_\text{N}$, each showing a different magnetic response.
In a ZF \musr\ experiment, 100\% spin-polarized muons are implanted into the material, stopping at interstitial sites where they precess in the local internal magnetic field arising from the sample's magnetism.
After an average of 2.2~$\mu$s, the muon decays, emitting a positron preferentially in the direction the muon spin was pointing at the time of decay.
Thus, by detecting positrons from many muon implantations, we can track the time evolution of the muon spin and understand the local magnetic field in the sample.

\begin{figure}
	\centering
	\includegraphics[width=0.99\linewidth]{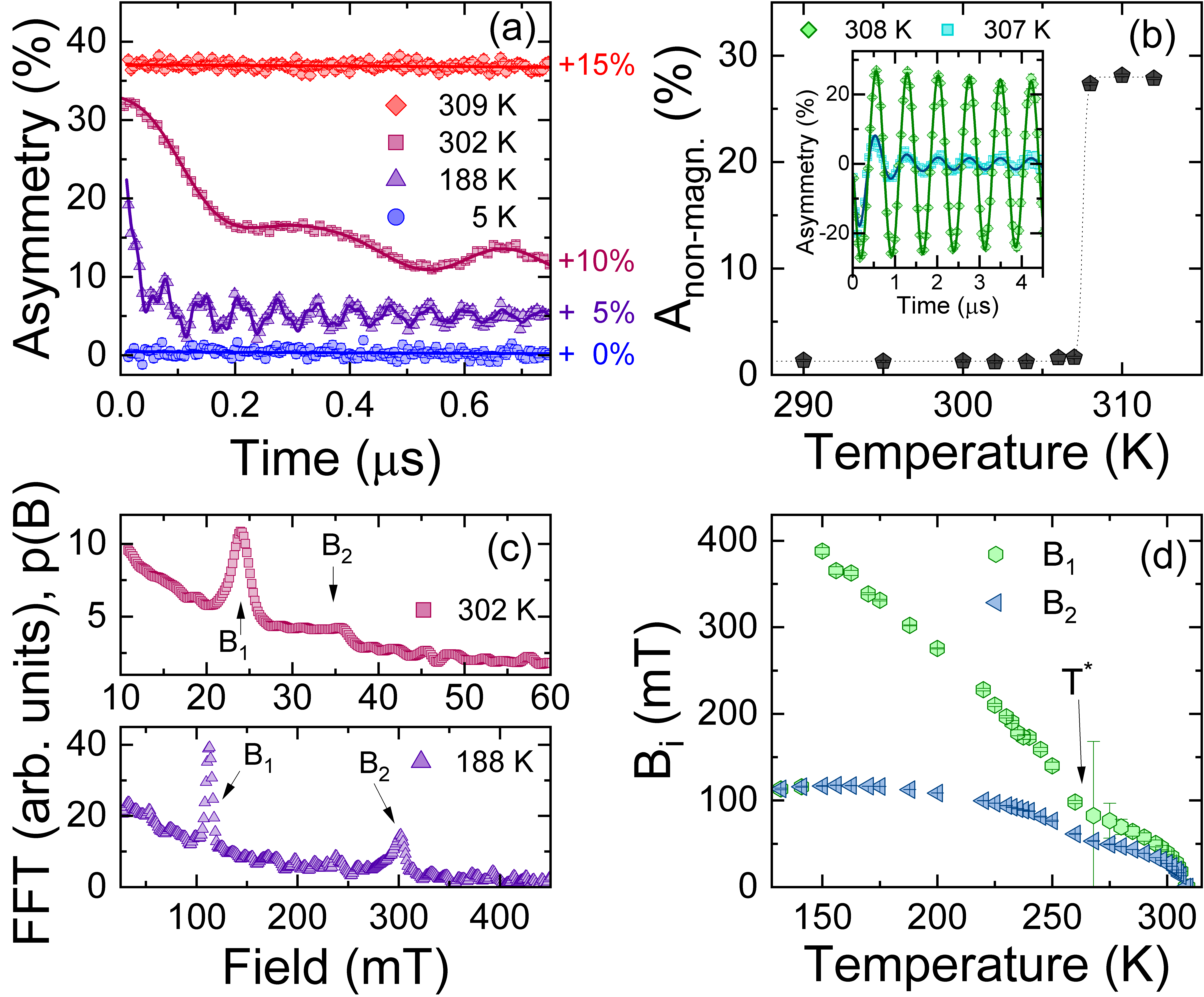}
	\caption{Muon-spin spectroscopy measurements of single-crystal MnTe. (a) Zero-field asymmetry spectra at selected temperatures, with marked vertical offsets. Coherent oscillations reveal long-range magnetic order. (b) Weak-transverse-field asymmetry in an applied field of 10~mT, which is proportional to the non-magnetic volume fraction. The inset shows corresponding time spectra above and below $T_\text{N}$. (c) Field distributions resulting from the Fast Fourier Transform (FFT) of the spectra in (a); two distinct internal fields are observed as peaks. (d) Internal fields $B_i$ at the muon stopping site, revealing a clear change in magnetic behavior around $T^*\simeq260$~K.}
	\label{fig:fig2}
\end{figure}

Representative ZF \musr\ asymmetry spectra (directly proportional to the muon spin polarization) are shown in Fig.~\ref{fig:fig2}(a).
Above $T_\text{N}=307.5(5)$~K, we observe a weak monotonic relaxation of the \musr\ asymmetry, typical of paramagnetic dynamics.
We confirm this observation by applying a weak transverse field (wTF) of 10~mT perpendicular to the initial muon spin direction, where only the spins of muons stopping in nonmagnetic regions precess in the applied field.
As shown in Fig.~\ref{fig:fig2}(b), fitting the data~\cite{supplement} yields an oscillating asymmetry of 26.6(6)\% above 308~K, consistent with the instrument's maximum asymmetry, thereby confirming the absence of long-range magnetic order above $T_\text{N}$.
Conversely, below $T_\text{N}$ there are two components in the wTF asymmetry: a fast-relaxing component from muons stopping inside the sample that are predominantly sensitive to the strong internal field of the long-range magnetic order, and a long-lived oscillating fraction attributed to muons stopping outside MnTe (accounting for 4.9(3)\% of the total measured muons, likely stopping in the sample holder or cryostat walls) that are subject only to the applied field.
The MnTe single crystal therefore exhibits long-range magnetic order in the entire sample volume for all $T<T_\text{N}$.
Similar results were obtained for other crystals (encompassing different growth methods), and a thin-film sample where only wTF measurements were possible, suggesting similar magnetism~\cite{supplement}.

To gain further insight into the magnetic behavior of MnTe, we perform a Fast Fourier Transform (FFT) of the ZF asymmetry spectra, yielding a measure of the internal field distributions [Fig.~\ref{fig:fig2}(c)].
As evident from the beating in the raw asymmetry spectra and the two peaks in the corresponding FFT, the muon ensemble is sensitive to two distinct internal fields over a wide temperature range, consistent with previous work over a more limited temperature range~\cite{yumnam2024magnon}.
By fitting the asymmetry spectra~\cite{supplement}, we track the temperature evolution of these two internal fields, denoted $B_{1,2}$ [Fig.~\ref{fig:fig2}(d)].
Two regimes of behavior emerge, separated by a clear kink at $T^*\simeq260$~K in the temperature evolution of $B_{1,2}$.
Evidence for this change is also present in the relaxation rates~\cite{supplement}, which peak at $T^*$; such behavior is typically only seen at magnetic phase transitions.
At higher $T$ ($T^*<T<T_\text{N}$), the ratio between the two internal fields is smaller ($B_1/B_2\simeq1\text{--}2$), whereas for $T<T^*$ this ratio increases with decreasing $T$ (reaching at least $B_1/B_2\simeq4$).
This abnormal temperature evolution of the internal fields suggests a complex evolution of the magnetic structure with temperature.

\begin{figure}
	\centering
	\includegraphics[width=\linewidth]{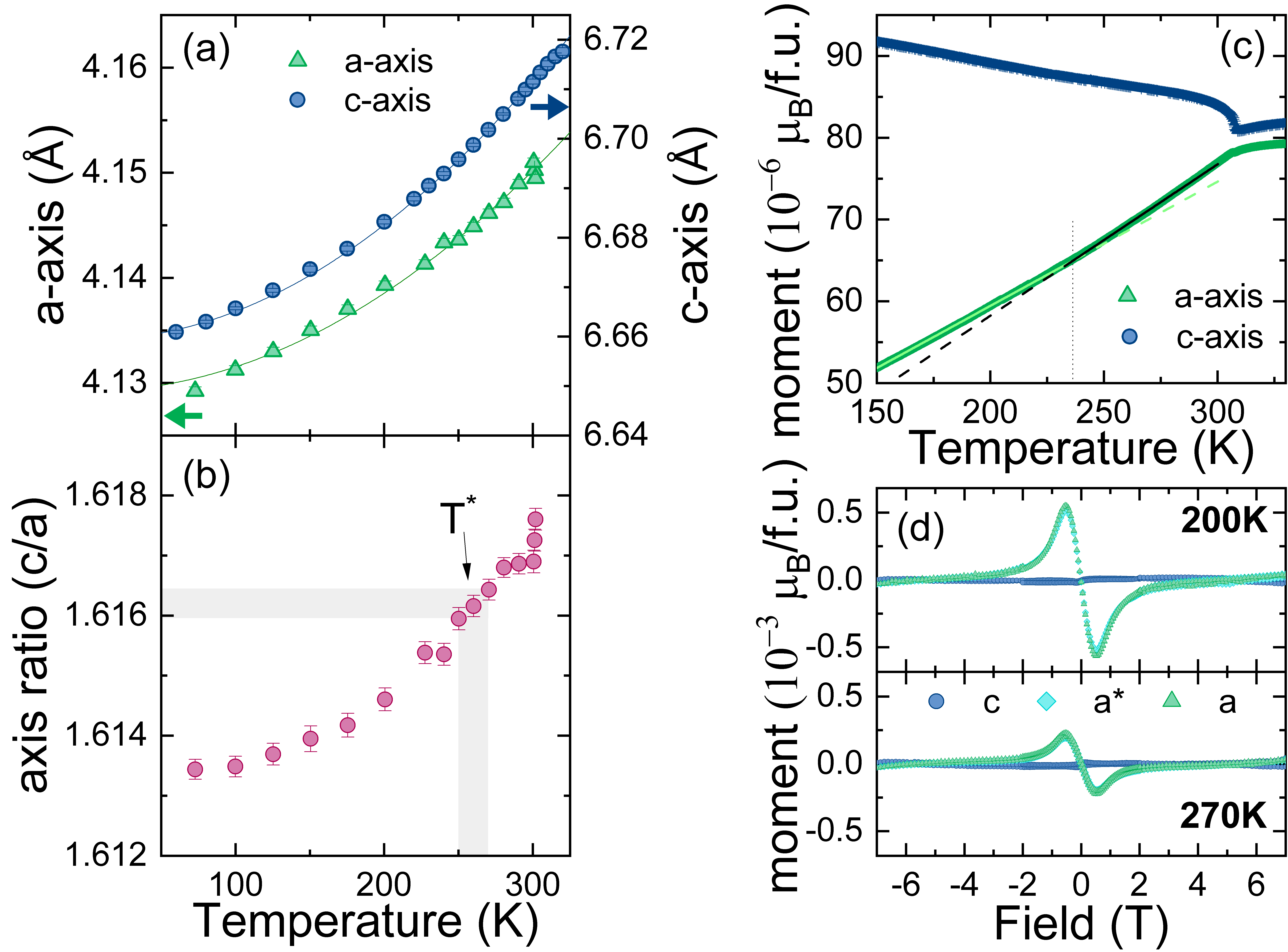}
	\caption{(a) Temperature dependence of the lattice parameters and (b) their ratio. The gray region highlights the axis ratio at $T^*$. (c) Temperature dependence of the magnetic moment in a 10~mT field (along $a$ or $c$) after field cooling. A small deviation from linearity is observed at the indicated temperature, near $T^*$. (d) Field dependence of the magnetic moment for different applied field directions (linear trend subtracted), highlighting the spin-flop transition for fields in the $ab$-plane. Similar results are obtained above and below $T^*$.}
	\label{fig:fig3}
\end{figure}

To understand this change in magnetic response at $T^*$, we turn to other experimental probes.
First, X-ray scattering extracts the lattice parameters of MnTe as a function of $T$ [Fig.~\ref{fig:fig3}(a)].
Both $a$ and $c$ increase smoothly and monotonically with $T$.
The unusual temperature dependence is phenomenologically well described by a $(1+\alpha T^2)$ behavior, consistent with neutron results~\cite{szuszkiewicz2005neutron}.
The $c/a$ ratio also increases with $T$, remaining slightly above 1.6 [Fig.~\ref{fig:fig3}(b)].
Magnetization measurements [Figs.~\ref{fig:fig3}(c--d)] are challenging due to the vanishing net magnetization; however, temperature-dependent measurements in an applied field of \SI{10}{mT} also show a weak change around $T^*$.
The measurement uncertainty remained constant (around $1\times10^{-9}$~emu) throughout the entire temperature scan, ruling it out as the cause of this slope change.
Measurements as a function of applied field along different crystallographic axes reveal differences between the $ab$-plane and $c$-axis [Fig.~\ref{fig:fig3}(c)].
Crucially, measurements above and below $T^*$ are unchanged for fields applied along $c$, whereas a notable change occurs for fields in the $ab$-plane.
This suggests that the transition at $T^*$ is predominantly associated with the in-plane orientation of the magnetic moments.

\begin{figure}
	\centering
	\includegraphics[width=0.95\linewidth]{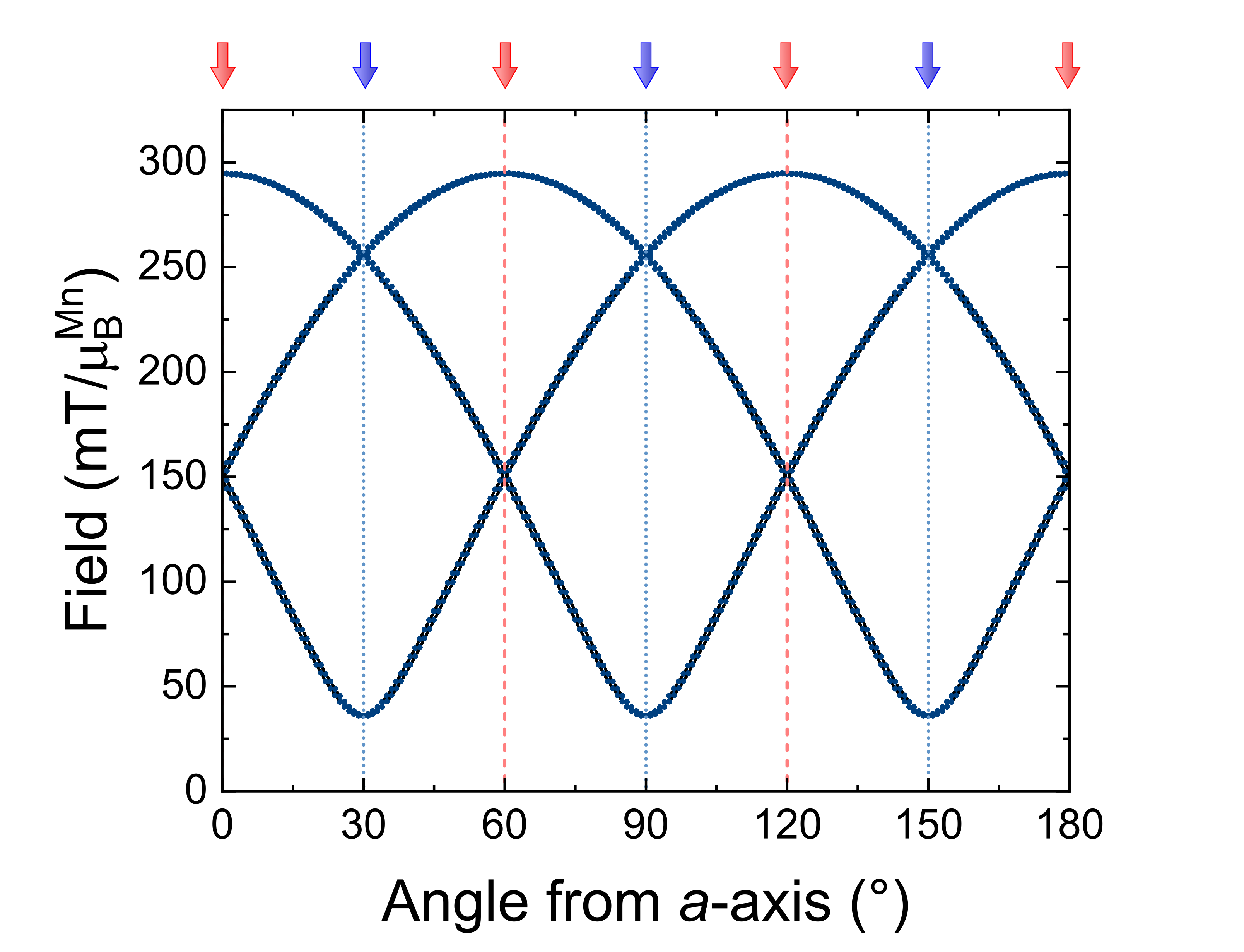}
	\caption{Calculated local magnetic field at the muon stopping site for different in-plane orientations of the Mn magnetic moments. Only two orientations (and their symmetric equivalents), indicated by arrows and dashed lines, yield the two internal fields observed experimentally.}
	\label{fig:fig4}
\end{figure}

To link these experimental observations, we performed muon stopping site calculations, identifying that the muon stops at the 6g $\sim(1/2, 0, 0)$ Wyckoff site.
We calculated the dipole field at this site assuming the magnetic structure of Refs.~\cite{kunitomi1964neutron,szuszkiewicz2005neutron,kriegner2017magnetic}, i.e., two opposite spin sublattices.
As previous work has struggled to unambiguously determine the magnetic easy axis, and the transition at $T^*$ likely involves the $ab$-plane, we evaluated the dipole field for various in-plane orientations of the magnetic moments.
The results are shown in Fig.~\ref{fig:fig4}.

As the magnetic moment rotates in the $ab$-plane, the number of distinct internal fields at the muon site changes.
Since we detect only two fields experimentally, the moments must point along one of two directions (and their symmetric equivalents): the $a$-axis, or 30$^\circ$ from the $a$-axis, which we denote as the $a^*$-axis.
All other directions would yield three measured internal fields.
For moments along the $a$-axis, the calculated ratio between the two internal fields is just under 2, matching our observations above $T^*$.
Conversely, for moments along the $a^*$-axis, the ratio is closer to 5, similar to what is observed at low $T$.
This indicates that below $T^*$, the magnetic moments point 30$^\circ$ from the $a$-axis, along $a^*$ (matching the previously proposed magnetic structure~\cite{kriegner2017magnetic}), before undergoing a spin reorientation transition at $T^*$ to align with the $a$-axis.

Interestingly, the ratio between the two fields is still not constant below $T^*$, necessitating a further, likely continuous, change to the magnetic structure.
A probable explanation is a temperature-dependent rotation or canting of the magnetic moments that yields a weak net magnetization, as suggested in Ref.~\cite{kluczyk2024coexistence}.
We find that several other subtle modifications to the magnetic structure can produce two internal fields (or fields clustering around two values) while significantly altering their ratio.

Although the observation of a second magnetic phase is perhaps surprising, several other techniques have detected changes around $T^*\simeq260$~K, supporting our findings.
In micrometer-sized MnTe structures, Ref.~\cite{amin2024nanoscale} reports an AHE and X-ray magnetic circular dichroism (XMCD) contrast at lower $T$ (150~K and 100~K, respectively), but both are absent at 250~K.
Both the XMCD signal~\cite{amin2024nanoscale,hariki2024x} and the AHE~\cite{gonzalez_betancourt_anisotropic_2024,gonzalez2023spontaneous} are expected to depend strongly on the angle of the altermagnetic order vector and vanish completely when the moments point along the $a$-direction~\cite{kluczyk2024coexistence}, consistent with our observations.
Other works~\cite{wasscher1965evidence,lee2025dichotomous,lee2024broken} report significant changes in angle-resolved photoemission spectroscopy and magnetotransport around $T\simeq260$~K.
While sometimes interpreted as the paramagnetic transition, this may be better explained by a change in the long-range magnetic order.

Neutron scattering, often the gold-standard probe for such changes in magnetic structure, struggles to distinguish between moments along $a$ and $a^*$.
Spin reorientations typically manifest as intensity anomalies in Bragg reflections upon passing through the transition temperature.
With equal magnetic domain populations, however, the structure factors produced by the $a$ and $a^*$ configurations are identical.
Our neutron diffraction measurements~\cite{supplement} reveal no anomalies at $T^*$, consistent with a 30$^\circ$ spin reorientation in the $ab$-plane assuming equally populated magnetic domains.

To understand the mechanism driving the spin reorientation transition at $T^*$, we turn to theory.
Extensive density functional theory (DFT) work on MnTe shows that while a Hubbard $U$ is essential to stabilize the correct magnetic structure~\cite{binci2025magnons}, spin-orbit coupling (SOC) has a negligible effect~\cite{chernov2025electronic}.
Furthermore, calculations of the magnetocrystalline anisotropy reveal very small energy differences for various in-plane moment orientations~\cite{kriegner2016multiple,kriegner2017magnetic,gonzalez_betancourt_anisotropic_2024}, supported experimentally by the ability to rotate the magnetic moments with an applied field~\cite{kriegner2016multiple}.
This naturally arises from the 3d$^5$ ($S=5/2$, $L=0$) orbital filling of Mn$^{2+}$ in MnTe; the half-filled shell quenches the orbital angular momentum, minimizing spin-orbit coupling, which normally dominates magnetic anisotropy.
Despite the complex magnetic exchange in MnTe~\cite{mazin2024origin}, we argue that the magnetic anisotropy is instead driven by crystal-field effects and single-ion anisotropies.

The effective Hamiltonian describing the anisotropy of a Mn moment in MnTe is expressed in terms of Stevens operators $O^m_n$ with only three nonzero coefficients $B_n^m$~\cite{hutchings1964point}: $B_2^0$, $B_4^0$, and $B_4^3$.
Assuming the magnetic moments lie at a small angle $\delta$ out of the $ab$-plane and an azimuthal angle $\phi$ from the $a$-axis, it can be shown~\cite{supplement} that
\begin{equation}
    \mathcal{H} \propto K_\text{out}\delta^2 - K_\text{in}\delta\cos\left(3\phi\right) ,
\end{equation}
where $K_\text{out}$ and $K_\text{in}$ are functions of $B_n^m$ that primarily control the out-of-plane and in-plane alignments, respectively.
Minimizing this Hamiltonian reveals that the moments do not lie perfectly in the $ab$-plane (i.e., $\delta\ne0$), but tilt by an angle
\begin{equation}
    \delta_\text{opt} = \frac{K_\text{in}\cos\left(3\phi\right)}{2K_\text{out}} ,
\end{equation}
yielding an angular-dependent part of $\mathcal{H}$
\begin{equation}
    \mathcal{H} \propto \frac{K_\text{in}^2\cos\left(6\phi\right)}{K_\text{out}} .
\end{equation}
This predicted out-of-plane tilt is consistent with other reports~\cite{kluczyk2024coexistence} and may explain the abnormal temperature dependence of the internal field ratio measured by \musr.
Notably, this out-of-plane tilt depends on the in-plane angle $\phi$.
Furthermore, the $\cos\left(6\phi\right)$ dependence dictates that for $K_\text{in}^2/K_\text{out} < 0$, the \neel\ vector aligns with the $a$-axis, whereas for $K_\text{in}^2/K_\text{out} > 0$, it points 30$^\circ$ away along the $a^*$-axis.

To explore whether a sign change in $K_\text{in}^2/K_\text{out}$ at $T^*$ drives the spin reorientation transition, we evaluate the $K_\text{in}$ and $K_\text{out}$ terms numerically.
These terms depend on $B_n^m$, which are functions of the temperature-dependent lattice parameters $a$ and $c$.
Neglecting the $B_4^0$ term, one finds~\cite{supplement} that $K_\text{in}^2/K_\text{out}\propto1/\left(3c^2-8a^2\right)$ for MnTe, with the sign of the proportionality constant remaining unchanged.
Thus, the \neel\ vector reorients at a critical ratio of $c/a=\sqrt{8/3}\simeq1.63$.
Including other terms (e.g., $B_4^0$) precludes a direct analytical evaluation of the critical ratio $c/a=\gamma_\text{crit}$, but does not fundamentally alter the physics.
From our experimental measurements [Fig.~\ref{fig:fig3}(b)], we determine $\gamma_\text{crit}\simeq1.616$, very close to the prediction of this simple model.

Previously reported values for the anisotropy in MnTe~\cite{szuszkiewicz2006spin,kriegner2017magnetic,liu2024chiral,dzian2025antiferromagnetic} vary between 20 and 250~$\mu$eV.
This large range likely stems from different sample geometries (thin films vs.\ bulk), diverse experimental and theoretical approaches, and the inherent difficulty of measuring such a small effect.
These values are consistent with our model, which expects small single-ion anisotropy.
Despite this small energy scale, the consequences are significant; Refs.~\cite{kriegner2016multiple,kriegner2017magnetic} note that the small in-plane anisotropy enables almost continuous variation of the resistance via the anisotropic magnetoresistance effect.
Additionally, because the lattice parameters remain very close to the critical value $\gamma_\text{crit}$ over the entire temperature range, small strains can switch the \neel\ vector at any $T<T_\text{N}$, as experimentally demonstrated in Ref.~\cite{liebman2026strain}.
Since strain is now suggested for preparing single-domain samples~\cite{liu2025strain}, care must be taken to account for the subtle magnetic changes it may induce.
MnTe provides a rare example of a material where small strains lead to significant changes in electrically measurable properties like the AHE, unlocking new research avenues and potential technological applications.
One such application is as a sensitive sensor of external stimuli: applying strain (an effective, albeit sample-dependent, tuning parameter for the AHE~\cite{liu2025strain,smolenski2025strain}) could produce an electrical switch-like response as the lattice parameters cross $\gamma_\text{crit}$.
This strain can be applied directly (producing a strain sensor) or coupled to other parameters (e.g., temperature via differential thermal expansion); because this effect is intrinsic to the material, miniaturization is unlikely to be a limiting factor.
Furthermore, the altermagnetic properties of MnTe (such as resilience to applied magnetic fields) and its significant magnetovolume effect~\cite{baral2023giant} further broaden its potential applications.

In summary, our \musr\ measurements of bulk MnTe reveal two distinct magnetically ordered regimes, explained by a change in the magnetic structure around $T^*\simeq260$~K.
Below $T^*$, the \neel\ vector lies 30$^\circ$ from the $a$-axis, before reorienting to align with the $a$-axis at higher $T$.
We attribute this transition to temperature-dependent single-ion anisotropies, likely driven by thermal expansion of the lattice parameters.
Additional, continuously evolving modifications to the magnetic structure may be needed to fully explain the \musr\ spectra; the small out-of-plane tilting of the Mn moments that arises naturally from our model is a likely candidate.
These discoveries highlight the extreme sensitivity of \musr\ to subtle magnetic changes, demonstrating its value for exploring altermagnets and other materials where precise magnetic details are critical.
Because the \neel\ vector reorientation has easily measurable consequences for all-electrical probes (such as the AHE or anisotropic magnetoresistance), this phenomenon opens new routes for fundamental research and potential technological applications.
As a room-temperature magnetically ordered material, MnTe is an exciting host for this effect, paving the way for device applications without the need for extensive materials engineering.
While this phenomenon may be rare because it requires orbital quenching, it is unlikely to be unique; theoretical work should therefore seek other candidate materials with similarly appealing combinations of structural parameters and orbitally quenched magnetic moments.

\section*{Acknowledgments}
Part of this work was carried out at \SmS\ and SINQ, \PSI, Switzerland; we are grateful for the provision of beamtime.
Samples were grown in MGML (mgml.eu), which is supported within the Program of Czech Research Infrastructures (Grant No. LM2023065).
KU and HR were supported by TERAFIT - CZ.02.01.01/00/22$\_$008/0004594 and the Dioscuri Program LV23025 funded by MPG and MEYS.
VM and ZM were supported by the Swiss National Science Foundation (Grant No.~200021-219950).
OC and MD were supported by the QM4ST project (CZ.02.01.01/00/22\_008/0004572) by MEYS CR. 
We wish to thank G.~Janka, T.~Prokscha, G.~Springholz,  J.~A.~Brock, A.~Hrabec, and A.~Suter for fruitful discussions, and J.~Volny for support with crystal growth.
AI tools (Gemini 3.1~Pro and Claude Opus~4.8) were used to check the manuscript for errors; suggestions were considered and, if appropriate, implemented by authors.

Early measurements were suggested by JHD, ZS, and JK.
The project was subsequently developed by TJH and JAK with input from many authors.
Single crystal samples were synthesised and characterised by KU, and DK.
Thin film samples were grown by OA, ADD, and PW.
Muon spin spectroscopy measurements were performed by TJH, HL, ZS, and JAK, and subsequently analysed by TJH and JAK.
magnetization measurements were performed by KU, HR, and JM.
MD and OC performed and analysed the temperature dependent X-ray diffraction. 
DFT calculations of the muon stopping site, and subsequent dipole field calculations were performed by TJH and JAK.
Neutron scattering measurements were performed by VCM, ZM, and OZ.
The theory linking the \neel\ vector and the crystal lattice parameters was developed by TJH.
The manuscript was initially written by TJH and JAK, and all authors discussed the results and refined the manuscript.

The authors declare no competing interests.

\section*{Data availability}
The data of this study will be made available before publication.

\section*{Appendix: Experimental and computational details}
Single crystals of MnTe were grown via the Bridgman technique using high-purity manganese (Mn, 5N8) and tellurium (Te, 6N).
Stoichiometric amounts of Mn and Te were loaded into a pointed-bottom alumina crucible and sealed under vacuum ($10^{-6}$~mbar) in a fused-silica ampoule.
The ampoule was heated in a vertical tube furnace at a rate of 1~$^\circ\text{C}$/min to 1150~$^\circ\text{C}$ and held at this temperature for 10 hours.
It was then slowly cooled to 900~$^\circ\text{C}$ at 0.03~$^\circ\text{C}$/min, after which the furnace was switched off.
The temperature gradient in the furnace was approximately 0.6~$^\circ\text{C}$/cm.
This yielded a single-crystalline ingot approximately 8~mm in diameter and 3~cm long, from which a 1~mm-thick slice was cut.

The \musr\ measurements were performed on the GPS spectrometer~\cite{amato2017new} at \SmS, \PSI, Switzerland.
The single crystal was oriented with the $a$-axis along the muon momentum direction (aligned with the forward-backward detectors), and the muon spin was rotated toward the $c$-axis (aligned with the up-down detectors) for ZF measurements.
In this geometry, the measurements are simultaneously sensitive to two components of the internal field.
To simulate these results, muon stopping sites were calculated using the MuFinder application~\cite{huddart2022mufinder}, followed by dipole field calculations with the muesr code~\cite{bonfa2018introduction}.

X-ray diffraction measurements were performed using a Rigaku SmartLab diffractometer equipped with a two-bounce Ge(220) monochromator operating at the Cu $K\alpha$ line.
The experiment was conducted in a custom-made helium cryostat (Cold Edge Technologies).
Symmetric $\theta$-$2\theta$ scans were collected at each temperature, and the lattice parameters were determined using all available diffraction peaks in the $2\theta$ range up to 140$^\circ$.
To extract both the $a$ and $c$ lattice parameters, measurements were conducted in two sample orientations: (0001) and (10$\bar{1}$0).

Magnetometry measurements were performed using a Quantum Design MPMS3 equipped with a quartz sample holder.
Single crystals were mounted with a small amount of nonmagnetic epoxy in various orientations relative to the applied magnetic field.

Density functional theory (DFT) calculations of the muon stopping sites were performed using the CASTEP code~\cite{clark2005first}.
The generalized gradient approximation for solids (PBEsol)~\cite{perdew2008restoring} was used throughout, with a Hubbard $U$~\cite{himmetoglu2014hubbard} of 4~eV applied to the Mn $d$ orbitals.
Crucially, spin-polarized calculations were employed to ensure the system maintained a band gap (as it does experimentally); without spin polarization, the calculations often converged to a metallic ground state, yielding incorrect muon stopping site energies.
A $3\times3\times2$ supercell was used to prevent self-interaction between muons (modeled as rescaled H atoms).
The lattice parameters were fixed to values obtained from geometry optimization of the pristine cell.
Calculations converged to better than 2~meV/atom using an 800~eV plane-wave cutoff and a $2\times2\times2$ $k$-point grid~\cite{monkhorst1976special}.
Atomic positions were subsequently relaxed until the energy and positions converged to better than $1\times10^{-6}$~eV/atom and $1\times10^{-4}$~\AA, respectively, with the maximum force on any atom below $2\times10^{-3}$~eV/\AA.

%\bibliography{Biblio}

%%%%%%%%%% Merge with supplemental materials %%%%%%%%%%
%% see https://tex.stackexchange.com/questions/168169/options-for-supplementary-
%% materials-in-preprint-version-revtex-arxiv
\clearpage
\onecolumngrid

\begin{center}
\section*{ Supplemental material for ``Thermally-driven reorientation of the \neel\ vector in altermagnetic MnTe''}
\end{center}
\twocolumngrid
%%%%%%%%%% Prefix a "S" to all equations, figures, tables and reset the counter
\setcounter{equation}{0}
\newcounter{FiguresInMainText}
\setcounter{FiguresInMainText}{\value{figure}}
\setcounter{table}{0}
\makeatletter
\renewcommand{\theequation}{S\arabic{equation}}
\renewcommand{\thefigure}{S\the\numexpr\value{figure}-\value{FiguresInMainText}}
\renewcommand{\bibnumfmt}[1]{[S#1]}
\renewcommand{\citenumfont}[1]{S#1}

\section{Muon Spin Spectroscopy}
This section provides further details regarding the fitting and results of our \musr\ measurements.
The wTF \musr\ data were fitted using
\begin{equation}\label{eq:wTF}
    A_\text{wTF} = \sum_{i=1}^2 A_i\exp\left(-\Lambda_it\right)\cos\left(\gamma_\mu Bt+\phi\right) ,
\end{equation}
where $\gamma_\mu=2\pi\times135.5$~MHz/T is the muon gyromagnetic ratio.
The two components with amplitudes $A_i$ correspond to muons stopping in different environments: magnetically ordered regions and paramagnetic or diamagnetic regions.
Above $T_\text{N}$, only a single component is required, as the sample is magnetically disordered.
The fitted field $B\simeq10$~mT is very close to the applied field, as expected.
The phase $\phi\simeq86.5(1)^\circ$ or $-4.4(1)^\circ$ for the up-down and forward-backward detectors, respectively, matches the \SI{6}{\degree} spin rotation employed during the wTF measurements.
The relaxation rate $\Lambda_i$ reflects the local field distribution at the muon site; it is large ($\gtrsim\SI{2}{\micro s^{-1}}$) for muons experiencing long-range magnetic order in MnTe, and small ($\simeq\SI{0.01}{\micro s^{-1}}$) for muons stopping outside the sample or above $T_\text{N}$.
The asymmetry $A_i$ corresponding to this small relaxation rate is denoted as $A_{\mathrm{non-magn.}}$ in the figures.

Zero-field (ZF) \musr\ measurements focused on the up-down detector pair, where coherent oscillations were most clearly observed.
A slight hint of oscillations in the forward-backward detector pair suggests that the internal field at the muon site lies very close to the $ab$-plane.
The data were fitted using
\begin{align}\label{eq:ZF}
    A_\text{ZF} &= A_\text{rel}\exp\left(-\lambda_\text{rel}t\right)  \notag \\
    &+ A_\text{osc}\left[ \left(1-f_{\text{B}2}\right)\exp\left(-\Lambda_{\text{B}1}t\right)\cos\left(\gamma_\mu B_1t+\phi_{\text{B}1}\right)\right. \notag \\
    &\left.+ f_{\text{B}2}\exp\left(-\Lambda_{\text{B}2}t\right)\cos\left(\gamma_\mu B_2t+\phi_{\text{B}2}\right) \right]  \notag\\
    &+ A_\text{const},
\end{align}
where each component is defined in the main text or in this Supplemental Material.
Certain components were set to zero depending on the temperature regime.
The total asymmetry, $A_\text{tot} = A_\text{rel} + A_\text{osc}+ A_\text{const}$, was allowed to decrease at low $T$ as components became too fast to resolve.
The two phases were refined globally when these components were present, yielding abnormal values of $\phi_{\text{B}1}=-65(3)^\circ$ and $\phi_{\text{B}2}=-81(2)^\circ$, which suggests complex magnetism.
The fractional weight $f_{\text{B}2}=1/3$ was fixed, although global refinement yielded similar values.
At temperatures slightly above $T^*$, the $B_1$ component is difficult to refine due to its large depolarization rate.
This is reflected by the large fitting uncertainties on two of the $B_1$ data points.
To consistently describe the spectra, the fit was constrained such that the ratio $B_1/B_2$ increased monotonically with decreasing temperature between \SI{305}{K} and \SI{268}{K}.
All data were fitted using musrfit~\cite{suter2012musrfit}.

\begin{figure}
	\centering
	\includegraphics[width=\linewidth]{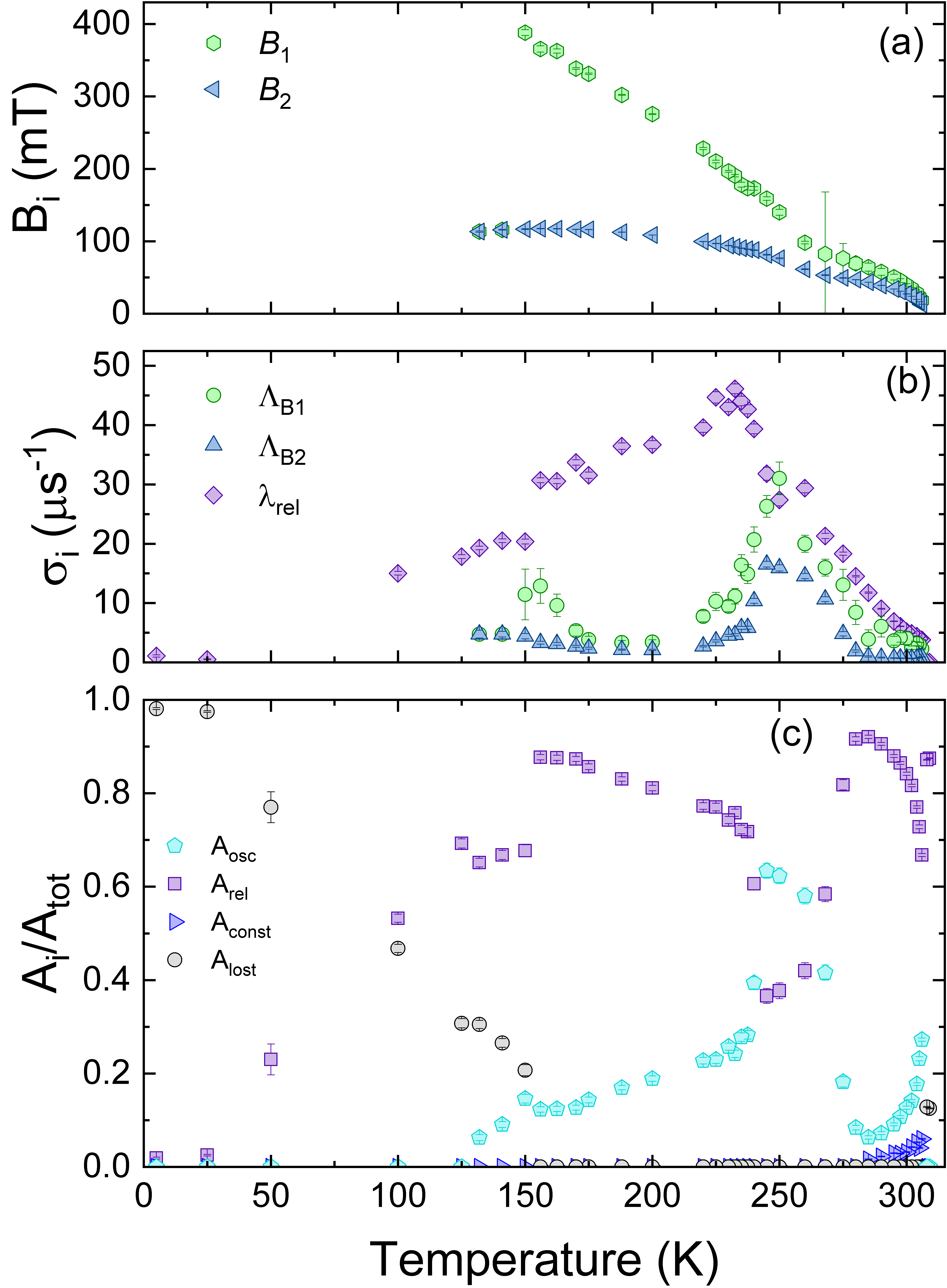}
	\caption{Temperature dependence of the ZF \musr\ fitting parameters. (a) Internal fields $B_i$, reproduced from the main text. (b) Corresponding relaxation rates $\Lambda_i$, alongside the fast early-time relaxation rate $\lambda_\text{rel}$. (c) Relative amplitudes of the various fit components.}
	\label{fig:figS1}
\end{figure}

The temperature-dependent fit parameters omitted from the main text are presented in Fig.~\ref{fig:figS1}.
Among these, the relaxation rates are particularly noteworthy.
The rate $\Lambda_i$, corresponding to the oscillating components, is primarily sensitive to the width of the static field distribution at the muon stopping site and is typically expected to scale with the internal field.
The fast relaxation rate $\lambda_\text{rel}$ characterizes muons experiencing a broad field distribution centered near zero.
Collectively, these relaxation rates provide a measure of static magnetic disorder.

As discussed in the main text, ZF \musr\ reveals the first magnetically ordered regime between $T^*\simeq260$~K and $T_\text{N}\simeq308$~K.
In this regime, a small time-independent term $A_\text{const}$ is present, which could arise from a small static field component along the $c$-axis or from a slight variation in the transition temperature across the sample.
The fast early-time relaxation described by $\lambda_\text{rel}$ suggests a significant disordered volume fraction, likely originating from magnetic domain walls.
Such domain walls, including vortex-like structures, have been observed previously in microscale MnTe structures~\cite{amin2024nanoscale}.
As the temperature decreases, the relaxation rates increase, peaking near $T^*\simeq260$~K.
This maximum indicates a region of peak disorder, as expected at a magnetic phase transition, providing further evidence for the spin reorientation.

We must also consider the possibility that the changing internal field ratio arises not from a change in the magnetic structure, but from macroscopic phase separation into distinct magnetically ordered domains.
However, the nearly temperature-independent amplitude ratio of the two oscillatory components strongly suggests that both originate from the same continuously evolving magnetic order.
Combined with the clear kink in the internal fields, which is difficult to reconcile with phase separation, and the aforementioned peak in the relaxation rates, we can confidently rule out macroscopic phase separation.

\begin{figure}
	\centering
	\includegraphics[width=0.6\linewidth]{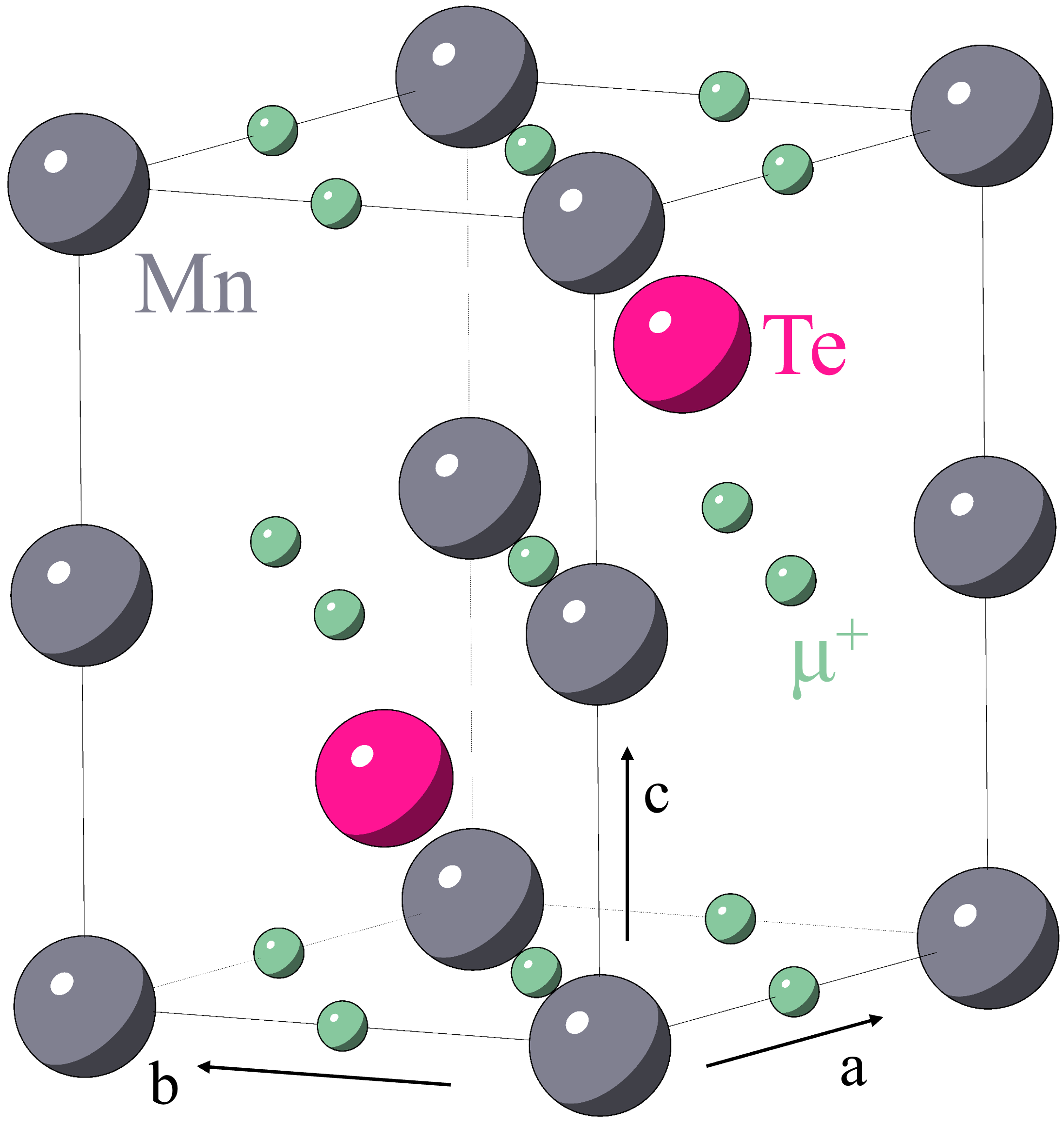}
	\caption{Calculated muon stopping sites in MnTe. The muon localizes at the $6g$ Wyckoff position. Slight structural distortions that must be accounted for in the dipole field calculations are too small to be visible here. Although only a single muon implants in the sample at a given time experimentally, all symmetrically equivalent positions are shown for completeness.}
	\label{fig:muonSite}
\end{figure}

\section{Muon stopping site and dipole field calculations}
Two candidate muon stopping sites were identified: one in the Te layer at the $6h$ ($x\simeq0.13$) Wyckoff position, and another in the Mn layer at the $6g$ position.
In both cases, the muon induces small local lattice distortions that must be accounted for in the dipole field calculations, though they are not large enough to compromise the muon's fidelity as a magnetic probe.
The calculated energy difference between the two sites is small, comparable to the computational variations arising from subtly different magnetic structures, precluding site assignment based solely on energetics.
However, symmetry dictates that only the $6g$ site yields the two distinct internal fields and the correct field orientation observed experimentally; we therefore conclude this is the sole occupied site [Fig.~\ref{fig:muonSite}].

To accurately simulate the dipole field at the muon stopping site, muon-induced crystallographic distortions were incorporated following Ref.~\cite{huddart2022mufinder}.
These distortions slightly cant the field out of the $ab$-plane (as the $c$-axis components no longer perfectly cancel), consistent with the small oscillating asymmetry observed in the forward-backward detector pair.
We assumed an equal population of all magnetic domains across all calculations.
While other reasonable assumptions (e.g., unequal domain populations or the presence of domain walls) would slightly modify the calculated dipole fields (particularly the ratio between the high and low fields and their relative weightings) these subtle variations do not alter our fundamental conclusions.

\begin{figure}
\centering
    \includegraphics[width=\linewidth]{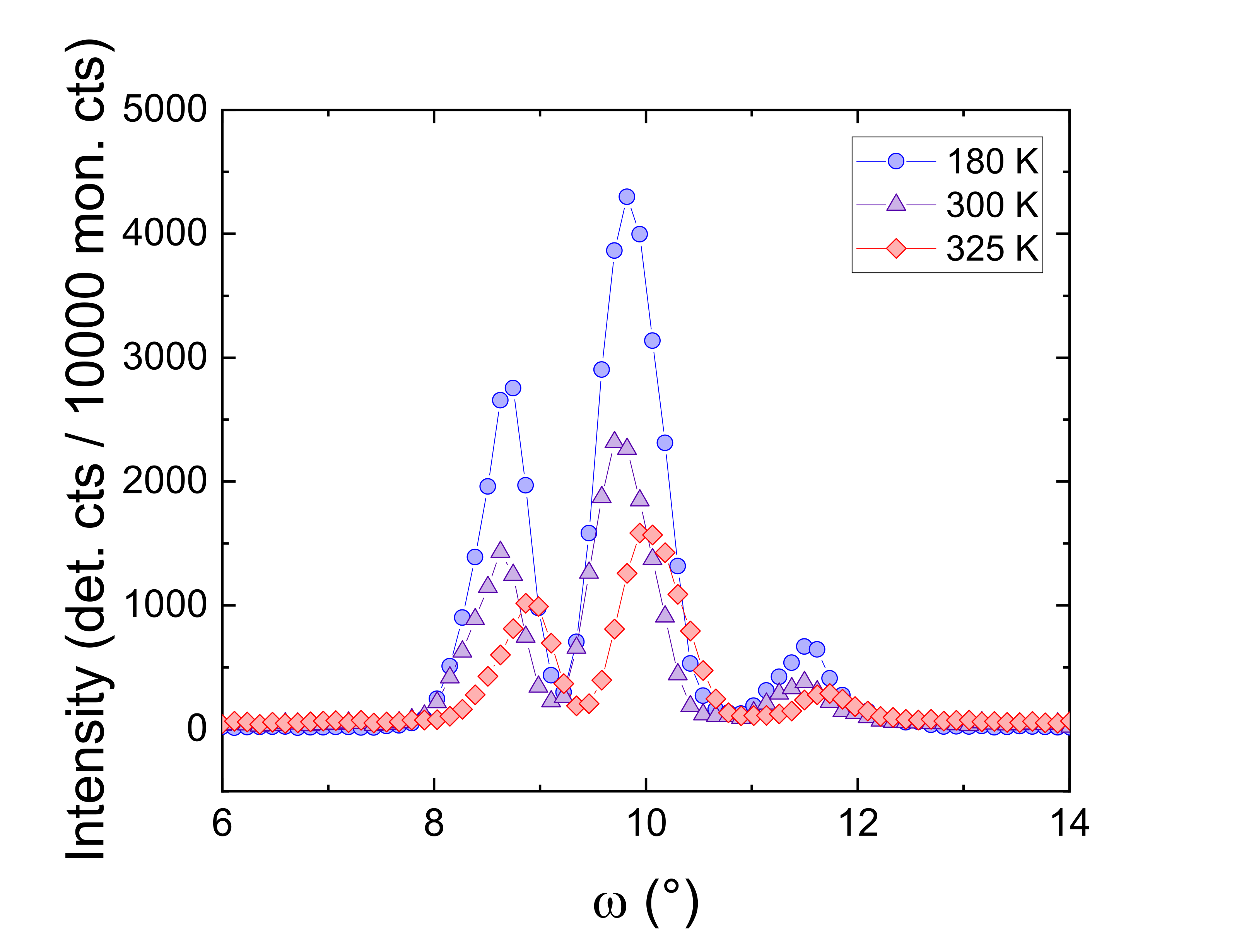}
    \caption{$\omega$-scan of the (011) reflection. The three peaks arise from distinct grains, each exhibiting a small shift in $\omega$ upon cooling from 325 to 300~K. Data were collected to a monitor count of $1\times10^4$ using a 198.2~mg piece of the sample discussed in the main text.}
    \label{fig:figS5}
\end{figure}

\section{Neutron scattering}
A 198.2~mg piece of the MnTe crystal discussed in the main text, consisting of approximately three grains with a mosaicity of $\sim3^\circ$ (Fig.~\ref{fig:figS5}), was mounted on an aluminum sample holder using a small amount of varnish.
The sample was oriented in the $(HHL)$ horizontal scattering plane, placed in a closed-cycle refrigerator, and measured on the Zebra thermal single-crystal diffractometer at the Paul Scherrer Institute.
Measurements were conducted in a 4-circle geometry using a wavelength of $\lambda = 1.178$~\AA\ (Ge-311 monochromator).
We collected 109 reflections at $T = 325$~K, 130 at 300~K, and 132 at 180~K.
Reflections with $2\theta\leq60^\circ$ were measured via $\omega$-scans, while those with $2\theta>60^\circ$ employed $\theta-2\theta$ scans.
Integrated intensities were extracted via trapezoidal integration following a linear background subtraction.
The 109 reflections collected at 325~K were used for nuclear refinement, while 123 and 125 reflections (those with $2\theta<67^\circ$) were used for magnetic refinements at 300~K and 180~K, respectively, using the FullProf software suite~\cite{rodriguez1993recent,rodriguez2025magnetic}.

The temperature and magnetic field dependencies of selected reflections were subsequently measured on Zebra in normal-beam geometry ($\lambda=1.178$~\AA) using 6~T and 10~T vertical-field cryomagnets.
For these experiments, the samples were similarly mounted on an aluminum holder and aligned in the $(HHL)$ plane, orienting the vertical field along the $[1\bar{1}0]$ ($\equiv \mathbf{a^*}$) direction.
The 198.2~mg crystal was used for the 6~T measurements, while a larger 279.25~mg portion of the same original crystal was used for the 10~T experiment.

\begin{figure}
\centering
    \includegraphics[width=\linewidth]{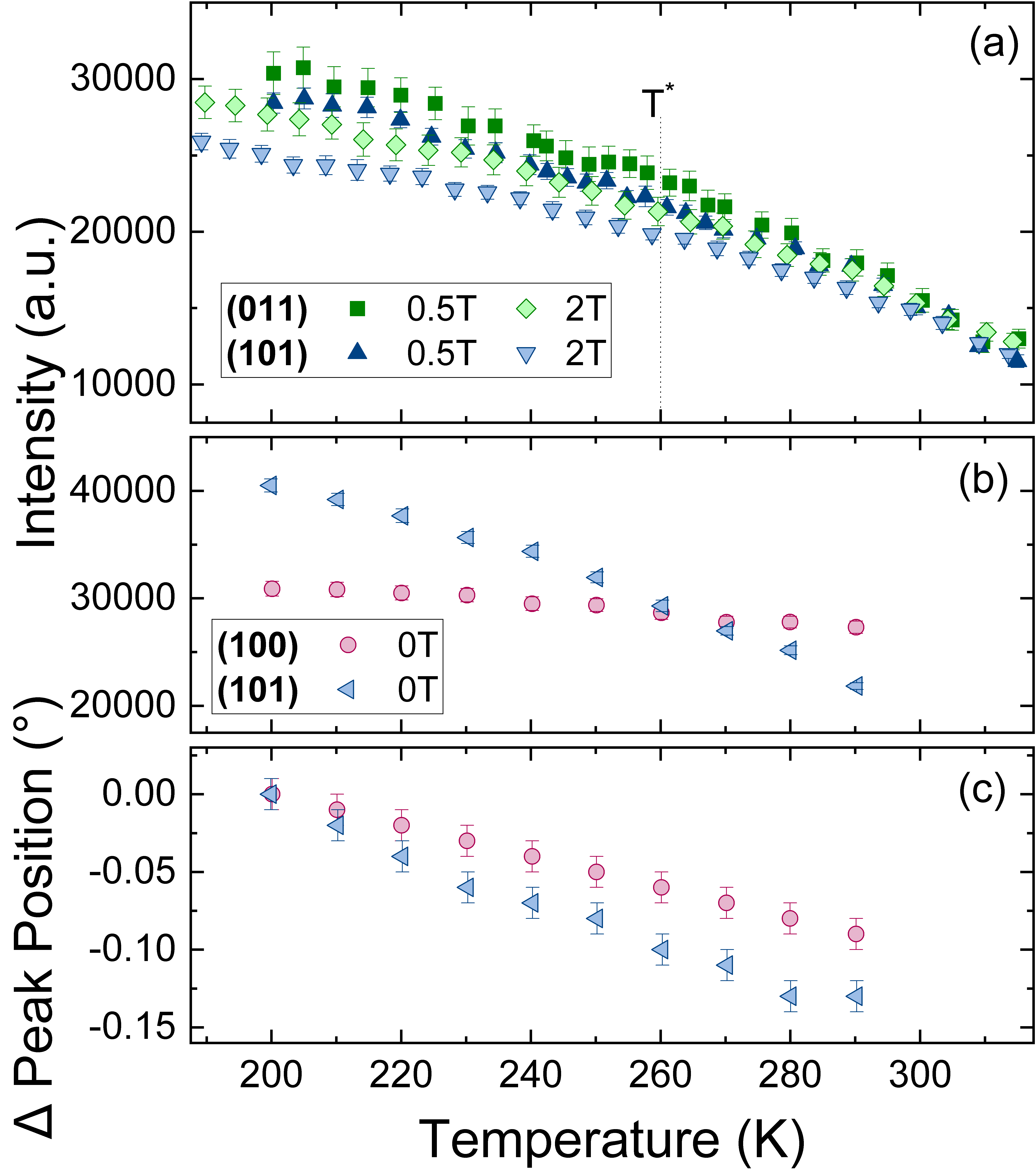}
    \caption{Temperature dependence of selected reflections. (a) Reflection intensities measured via $\omega$-scans upon warming, following field cooling at 0.5~T and 2~T. The intensities converge above 300~K. The dashed line denotes $T^*$. An intensity difference between the (101) and (011) reflections indicates a field-induced shift in the magnetic domain population. (b) Amplitudes and (c) peak positions of reflections measured via zero-field $\omega$-scans upon warming reveal smooth temperature evolution without anomalies. Although the (100) reflection exhibits a weak $T$-dependence, it lacks a distinct onset at $T^*$; such gradual variation is expected intrinsically due to the Debye-Waller factor.}
    \label{fig:figSneutronsT}
\end{figure}

Temperature scans from 200 to $\approx300$~K were obtained at 0, 0.5, and 2~T [Fig.~\ref{fig:figSneutronsT}] following field cooling from 315~K.
The intensity of the antiferromagnetic reflections increases upon cooling as the Mn$^{2+}$ moments order below $T_\text{N}$; no other anomalies are observed.
Reflections with in-plane components shift to lower $\omega$ upon warming, indicating an increase in the lattice parameter $a$, consistent with our X-ray diffraction results.
While no intensity anomalies are observed, as expected for the proposed in-plane reorientation of the magnetic easy axis, the onset of a staggered or net magnetization along the $c$-axis would produce a measurable anomaly in the Bragg intensities.
An alternative method to detect this reorientation is to introduce a domain imbalance via external stimuli (e.g., strain or a magnetic field).
For instance, a magnetic field applied along the $[1\bar{1}0]$ direction favors domains with moments most perpendicular to the field, allowing the (101) reflection to distinguish between the $a$ and $a^*$ configurations.
However, in addition to shifting the domain population, such an applied field also tilts the moments toward identical spin-flopped states for each configuration, obscuring the true zero-field magnetic ground state.
This underscores the necessity of a multi-technique approach and highlights the unique utility of \musr, which can detect this reorientation in zero field.

\begin{figure}
\centering
    \includegraphics[width=\linewidth]{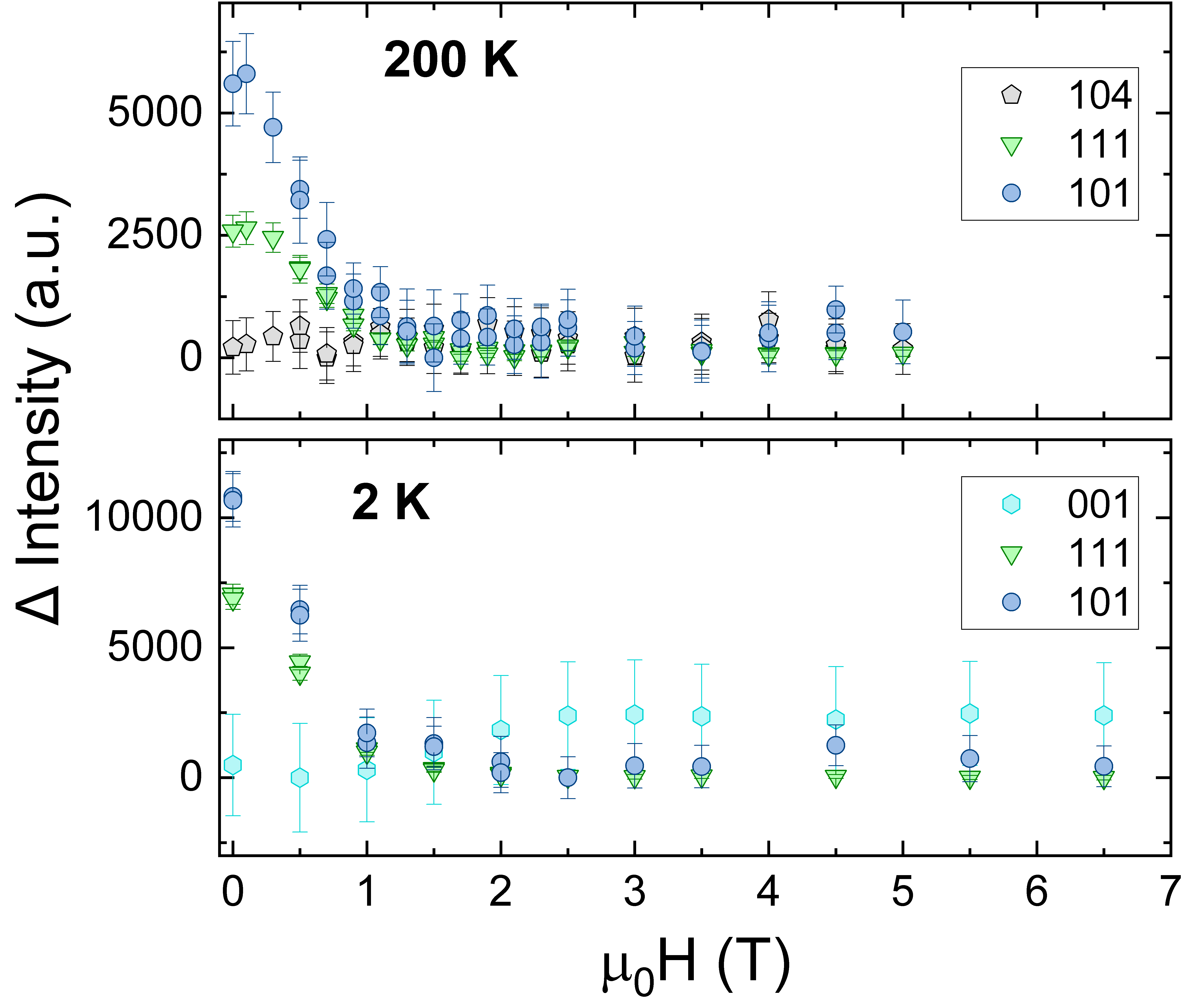}
    \caption{Magnetic field dependence of selected reflections. Intensities were measured via $\omega$-scans while sweeping the applied field up and down at 200~K (upper panel) and 2~K (lower panel). The plotted values represent the difference relative to the minimum intensity measured across the entire field sweep.}
    \label{fig:figS8}\label{fig:figSneutronsH}
\end{figure}

Magnetic field scans were performed at 2~K and 200~K [Fig.~\ref{fig:figSneutronsH}].
Reflections arising from the altermagnetic order begin to lose intensity by 0.3~T, signaling either a spin-flop transition or domain repopulation.
From 2~T up to the maximum applied field, the intensity remains constant, exhibiting no hysteresis between the increasing and decreasing field sweeps.

\begin{table*}
    \caption{Nuclear and magnetic neutron diffraction refinement parameters. The overall scale factor of 7800(200), determined at 325~K, was applied to the 300~K and 180~K refinements. ``Corr.'' lists refined parameters exhibiting greater than 50\% correlation. ``D$x$'' denotes the population of domain $x$. ``Fixed'' indicates that the domain populations were constrained to the values obtained at 180~K, while ``ED'' denotes an assumption of equal domain populations.}
    \label{tab:neutron}
    \begin{ruledtabular}
    \begin{tabular}{cccccccccc}
    $T$~(K) & Config. & $\chi_\nu^2$ & $R$ & $B_\text{iso}$~(\AA$^2$) & $\mu~(\mu_\text{B})$ & D1 & D2 & D3 & Corr. \\
    \colrule
    325 & Nuclear & 99.5 & 5.74 & 1.5(1) & - & - & - & - & Scale, $B_\text{iso}$: 86\% \\
    & & & & & & & & & \\
    300 & $\mathbf{a}$ & 123. & 5.70 & 1.36(4) & 1.72(9) & 0.3(2) & 0.2(2) & 0.4 & $\mu$, D2: 66\% \\
    & $\mathbf{a^*}$ & 123. & 5.70 & 1.36(4) & 1.71(5) & 0.3(2) & 0.4(2) & 0.3 & - \\
    & $\mathbf{a}$ Fixed & 123. & 5.75 & 1.37(4) & 1.81(7) & 0.30 & 0.43 & 0.27 & - \\
    & $\mathbf{a^*}$ Fixed & 122. & 5.74 & 1.36(4) & 1.70(4) & 0.37 & 0.24 & 0.39 & - \\
    & & & & & & & & & \\
    180 & $\mathbf{a}$ & 203. & 7.07 & 1.30(5) & 3.65(8) & 0.30(7) & 0.43(9) & 0.27 & $\mu$, D2: 69\% \\
    & $\mathbf{a^*}$ & 203. & 7.05 & 1.30(5) & 3.66(4) & 0.37(7) & 0.24(6) & 0.39 & - \\
    & $\mathbf{a}$ ED & 206. & 7.11 & 1.26(5) & 3.55(6) & 1/3 & 1/3 & 1/3 & - \\
    & $\mathbf{a^*}$ ED & 204. & 7.12 & 1.30(5) & 3.66(4) & 1/3 & 1/3 & 1/3 & - \\
    \end{tabular}
    \end{ruledtabular}
\end{table*}

A nuclear refinement of the 325~K dataset (``Nuclear'' in Table~\ref{tab:neutron}) was performed to determine the overall scale factor for subsequent magnetic refinements.
The isotropic thermal parameter $B_\text{iso}$ was also refined, exhibiting a strong correlation with the scale factor.

\begin{figure}
\centering
    \includegraphics[width=\linewidth]{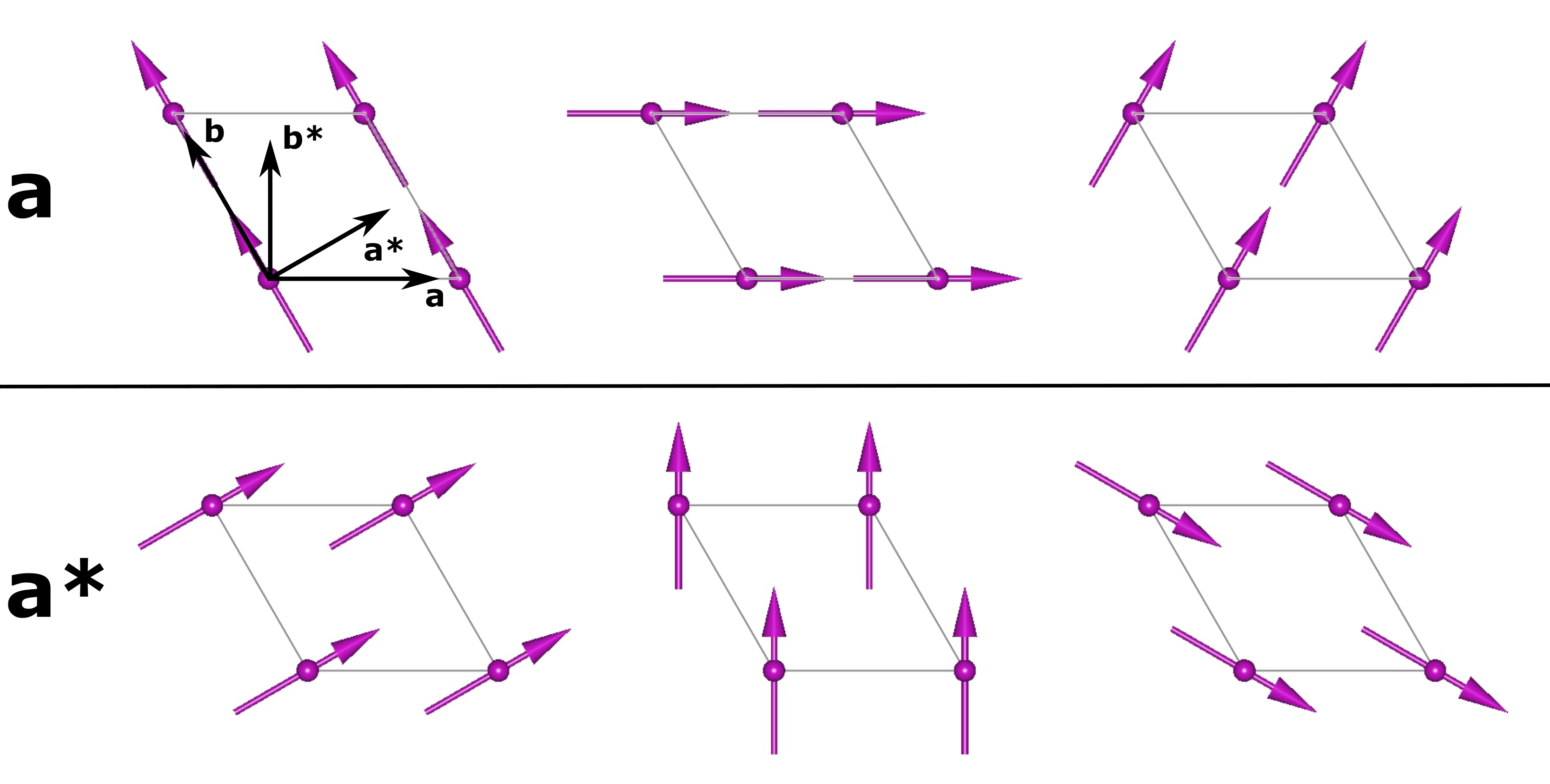}
    \caption{Possible magnetic domains in MnTe. The three distinct domains are shown for the $\mathbf{a}$ (top) and $\mathbf{a^*}$ (bottom) configurations.}
    \label{fig:figS9}
\end{figure}

Magnetic refinements were performed in the spin-compensated ordered state at 300~K.
Because the propagation vector is $\mathbf{k} = (000)$, the magnetic and nuclear reflections coincide; therefore, both nuclear and magnetic phases were included in the refinements.
The magnetic moments were constrained to lie along either the $\mathbf{a}$ or $\mathbf{a^*}$ direction.
Domain averaging over moment orientations was included to account for the other two symmetrically equivalent domains in zero field (Fig.~\ref{fig:figS9}).
With the scale factor fixed to the value determined at 325~K, the isotropic thermal parameter $B_\text{iso}$, the magnetic moment $\mu$, and the domain populations were refined (rows ``$\mathbf{a}$'' and ``$\mathbf{a^*}$'' in Table~\ref{tab:neutron}).
Additionally, $B_\text{iso}$ and $\mu$ were refined with the domain populations fixed to their corresponding values obtained at 180~K (rows ``$\mathbf{a}$ Fixed'' and ``$\mathbf{a^*}$ Fixed'' in Table~\ref{tab:neutron}).
Although the experiment is sensitive to the magnetic moment $\mu$, it cannot distinguish between the $\mathbf{a}$ and $\mathbf{a^*}$ configurations; despite differences in the refinement minima, both configurations yield identical magnetic structure factors.
While neutron diffraction is, in principle, sensitive to the orientational domain populations, the presence of multiple grains and the large mosaicity of the crystal precluded a reliable refinement of these populations at 300~K.

Magnetic refinements at 180~K were also performed with the scale factor fixed to the 325~K value.
Initially, $B_\text{iso}$, $\mu$, and the magnetic domain populations were co-refined (rows ``$\mathbf{a}$'' and ``$\mathbf{a^*}$'' in Table~\ref{tab:neutron}).
The resulting domain populations are inverted between the two configurations: in the ``$\mathbf{a}$'' configuration, domains 1 and 3 are less populated than domain 2, whereas in the ``$\mathbf{a^*}$'' configuration, domain 2 is less populated than domains 1 and 3.
A second set of refinements was performed with the domains constrained to equal populations (rows ``$\mathbf{a}$ ED'' and ``$\mathbf{a^*}$ ED'' in Table~\ref{tab:neutron}).
Across all four of these 180~K refinements, the magnetic moment increases relative to the 300~K results, while remaining below the saturated value of $5~\mu_\text{B}$, and the thermal parameter $B_\text{iso}$ slightly decreases.

\section{M\lowercase{n}T\lowercase{e}$_2$ impurity}
The sample discussed throughout the main text (sample A) was grown via the Bridgman technique.
The samples primarily discussed below in this Supplemental Material (samples B and C) were grown using a Te-rich self-flux method, following the procedure detailed in Ref.~\cite{krempasky2024altermagnetic}.
This yielded high-quality platelike single crystals with lateral dimensions of several millimeters and thicknesses of several hundred micrometers.
The crystallographic $c$ axis is oriented normal to the plate surface.

Scanning electron microscopy (SEM) revealed MnTe$_2$ impurities in all samples.
In the Bridgman-grown crystal (sample A), MnTe$_2$ appeared as precipitates aligned along the (0001) planes within the bulk, constituting a volume fraction below 2\%.
Conversely, in the flux-grown crystals (samples B and C), MnTe$_2$ formed exclusively on the surface and was subsequently removed via mechanical polishing.

\begin{figure}
	\centering
	\includegraphics[width=\linewidth]{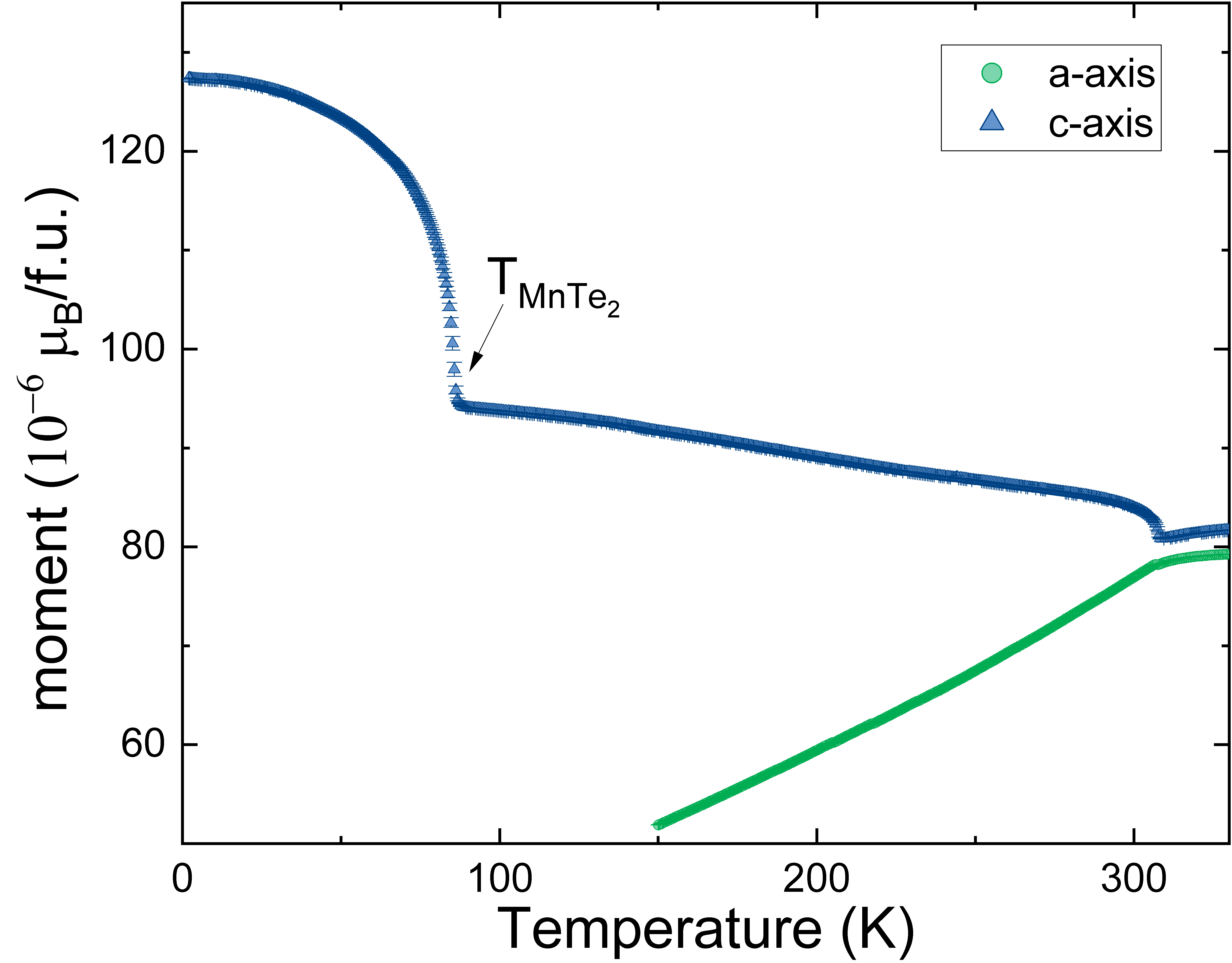}
	\caption{Magnetization of MnTe containing MnTe$_2$ impurities. Field-cooled magnetization measured at 10~mT as a function of temperature. The impurity phase produces a prominent anomaly near 80~K. Data were collected from sample A, the crystal discussed in the main text.}
	\label{fig:figS3}
\end{figure}

Figure~\ref{fig:figS3} displays the magnetization of the MnTe crystal discussed in the main text, which contains a minor MnTe$_2$ impurity phase.
The MnTe$_2$ magnetic ordering transition is clearly visible near 80~K; despite occupying a small volume fraction of the sample, the impurity yields a comparatively large magnetization.
This points to a ferromagnetic response from the impurity, generating significant random stray fields that affect muons implanting in the MnTe bulk.
These stray fields account for the loss of asymmetry in our \musr\ measurements, as they rapidly dephase the muon spin polarization and preclude reliable data collection in this temperature regime.

Previous studies have demonstrated that the domain population, and consequently the magnetization, in MnTe is highly sensitive to both magnetic field history~\cite{rossler2025low} and relatively weak applied fields, even along the hard $c$-axis~\cite{watanabe2026magneto}.
Consequently, interpreting bulk magnetometry data is challenging, particularly when attempting to resolve subtle phenomena such as the reorientation of the \neel\ vector.
This further underscores the distinct advantage of local probes like \musr, which can evaluate these effects under true zero-field conditions.

In the measurements presented in the main text, the initial asymmetry decreases at low $T$, eventually becoming unresolvable.
This indicates that the internal fields or relaxation rates, which reflect the width of the local magnetic field distribution, become too large for the time resolution of the \musr\ spectrometer, causing the muon polarization to dephase rapidly.
The onset temperature of this signal loss correlates with the ferromagnetic ordering temperature of MnTe$_2$~\cite{sawaoka1966effect}, which forms inclusions throughout the bulk sample A.
Furthermore, theoretical studies suggest that strain can shift the ordering temperature of MnTe$_2$ by up to a factor of three~\cite{chen2024large}; because the MnTe$_2$ inclusions are likely strained by the surrounding MnTe matrix, the progressive loss of initial asymmetry probably arises from a gradual, spatially distributed ferromagnetic ordering of these impurity phases.

\begin{figure*}
	\centering
	\includegraphics[width=0.8\linewidth]{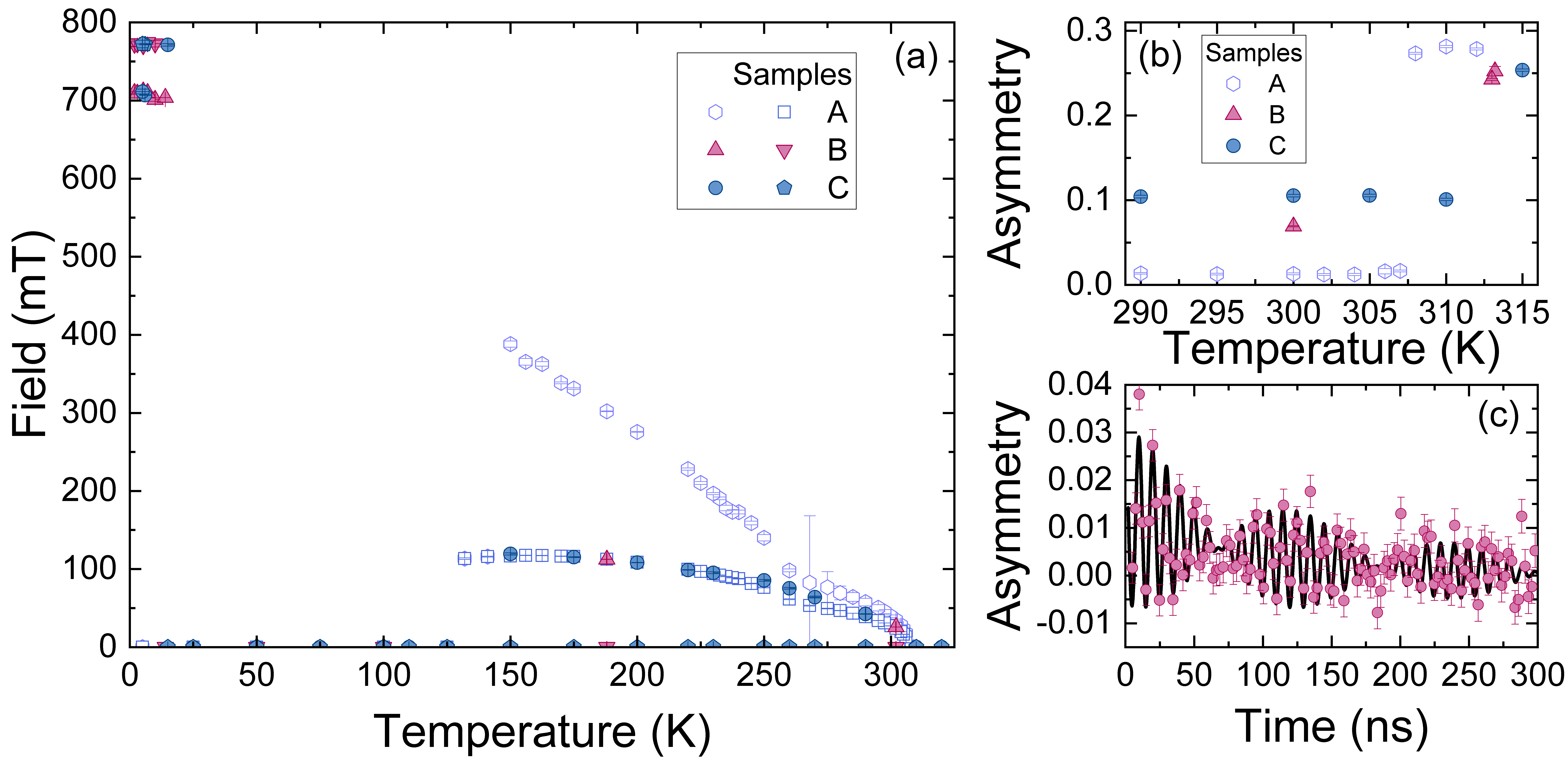}
	\caption{\musr\ measurements on MnTe$_2$-free samples. (a) Temperature dependence of the ZF-\musr\ internal fields across different samples. Points plotted at zero indicate unresolvable frequencies. (b) Weak transverse-field asymmetry in these samples near the transition temperature. (c) Sum of all ZF-\musr\ spectra collected below 10~K for sample B.}
	\label{fig:figS2}
\end{figure*}

To verify the intrinsic low-$T$ behavior of MnTe, we performed zero-field (ZF) \musr\ measurements on the two small, flux-grown MnTe crystals (B and C) free of MnTe$_2$ impurities.
Given their shape and size, these crystals were mounted with the $c$-axis parallel to the beam direction, resulting in oscillations primarily observed in the forward-backward detector pair.
The small sample volume introduces significant experimental challenges, notably a large background asymmetry from muons implanting in the silver degrader, sample holder, or cryostat walls.
Consequently, the oscillating asymmetry in a weak transverse field does not drop to zero below $T_\text{N}$ [Fig.~\ref{fig:figS2}(b)], and only the higher-amplitude ZF oscillation frequency is resolvable above 150~K.

Despite these challenges, the ZF oscillation frequencies [Fig.~\ref{fig:figS2}(a)] closely match those of the larger crystal, confirming that sample A remains a faithful probe of the intrinsic MnTe magnetism at higher temperatures.
Crucially, oscillations re-emerge in the MnTe$_2$-free samples below 15~K, exhibiting a beating pattern in the time spectra indicative of two closely spaced high frequencies [Fig.~\ref{fig:figS2}(c), representing $\sim19$~h of integrated acquisition below 10~K for sample B].
The magnitude of these internal fields is consistent with a low-temperature extrapolation of $B_1$.
This $\simeq10\%$ splitting of the $B_1$ field suggests the magnetic moments cant slightly away from the $a^*$ axis (as supported by the muon site calculations in Fig.~4 of the main text).
Such a subtle temperature dependence of the \neel\ vector accounts for the evolving $B_1/B_2$ ratio below $T^*$.
Additionally, an anomaly in the low-temperature resistivity of similarly grown MnTe crystals~\cite{dzian2025antiferromagnetic} further corroborates a minor structural or magnetic adjustment.
Ultimately, we conclude that the fundamental magnetic structure of MnTe does not undergo significant changes below $T\simeq100$~K, aside from this slight canting of the \neel\ vector.

\section{Low energy muon measurements on a M\lowercase{n}T\lowercase{e} film}

\begin{figure*}
	\centering
	\includegraphics[width=0.8\linewidth]{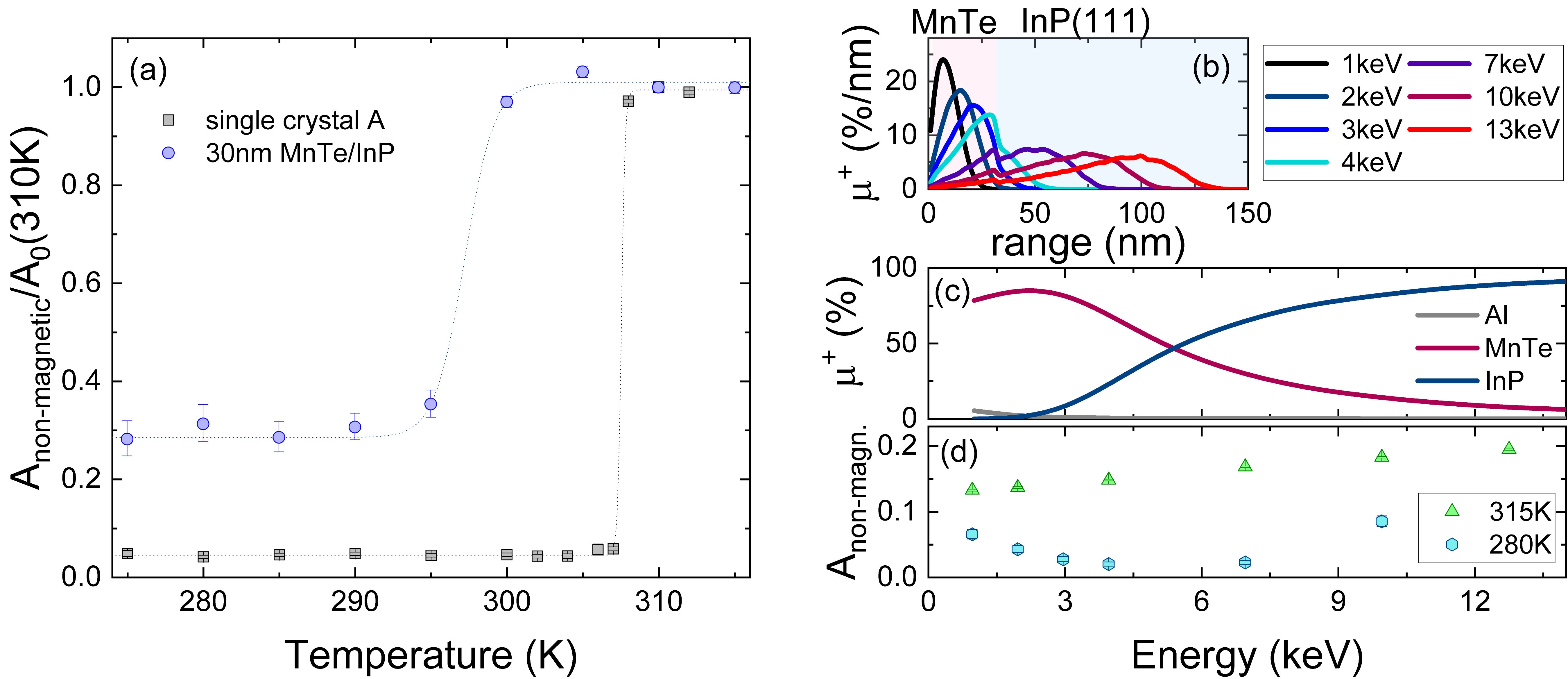}
	\caption{Low-energy muon-spin spectroscopy. (a) Nonmagnetic component of the weak transverse-field asymmetry (10~mT applied field) for 2~keV muons implanted in a MnTe thin film, compared with the bulk measurements of sample A from the main text. Dotted lines are guides to the eye highlighting the magnetic transition. (b) Simulated muon implantation profiles at various incident kinetic energies. (c) Corresponding muon stopping fraction per layer. (d) Implantation energy dependence of the 10~mT weak transverse-field asymmetry above and below the magnetic transition.}
	\label{fig:figSlem}
\end{figure*}

To complement the bulk MnTe results, low-energy \musr\ was performed on a $15 \times 20$~mm$^2$ film consisting of 2~nm Al / 30~nm $\alpha$-MnTe.
The film was grown at $\sim700$~K via molecular beam epitaxy (MBE) on an InP(111)A substrate, with the MnTe $c$-axis oriented normal to the substrate surface.
We confirmed the correct crystallographic phase and growth orientation using x-ray diffraction. 
Measurements were conducted using the LEM spectrometer~\cite{morenzoni2000lem,salman2012design} at the $\mu$E4 beamline~\cite{prokscha2008muE4} of the Swiss Muon Source at the Paul Scherrer Institut, Switzerland.
Low-energy muons were generated in a solid-N$_2$-capped solid-Ar moderator and electrostatically guided to the sample.
The film was affixed to a Ni-coated sample plate using silver paint and placed in ultrahigh vacuum, where the temperature was regulated by a cold-finger $^4$He cryostat.
The initial muon spin polarization was directed in-plane, and the implantation energy was tuned via a high-voltage bias applied to the sample plate.

A clear magnetic transition is observed in a 10~mT weak transverse field applied out-of-plane [Fig.~\ref{fig:figSlem}(a)].
The 2~keV implantation energy used for this measurement centers the muon stopping distribution approximately within the middle of the MnTe layer, as confirmed by \texttt{Trim.SP}~\cite{morenzoni2002implantation} simulated stopping profiles [Fig.~\ref{fig:figSlem}(b)].
The magnetic transition in the thin film occurs at a slightly lower temperature ($T_\text{c} \sim 297(2)$~K) than in the bulk.
Comparing the energy dependence of the weak transverse-field asymmetry with the expected layer-by-layer stopping fractions [Fig.~\ref{fig:figSlem}(c,d)] demonstrates that the entire MnTe layer becomes fully magnetic below the transition. 
While signatures of this transition are also evident in the zero-field spectra, we could not resolve coherent zero-field oscillations in the thin film.
This absence is likely due to the reduced time resolution inherent to the LEM instrument, which arises from time-of-flight broadening of the muon arrival times.

\section{Derivation of link between \neel\ vector and lattice parameters}
This section details the full derivation linking the \neel\ vector to the lattice parameters of MnTe, as introduced in the main text.
Although this effective model describes the experimental data well, it does not explicitly incorporate phenomena such as anisotropic exchange coupling or the semiconducting nature of MnTe.
Nevertheless, we expect these omitted effects to primarily modify the quantitative results, rather than undermine the fundamental conceptual framework established by our model.

\subsection*{Single ion anisotropy}
For an ion with a half-filled $d$-shell, such as Mn$^{2+}$ in MnTe ($3d^5$, $S=5/2$, $L=0$), the orbital angular momentum is quenched.
Consequently, one might expect an absence of spin-orbit coupling and, by extension, zero magnetic anisotropy.
Since this is experimentally not the case, the observed anisotropy must originate from higher-order perturbative effects, specifically crystal-field splitting and single-ion anisotropy.

The effective Hamiltonian describing the spin anisotropy can be expressed in terms of the Stevens operators $O^m_n$.
For a site with hexagonal symmetry (such as Mn in MnTe), the only terms with nonzero coefficients $B_n^m$ are $B_2^0$, $B_4^0$, $B_4^3$, $B_6^0$, $B_6^3$, and $B_6^6$~\cite{hutchings1964point}.
Furthermore, Stevens operators require $n\leq2S$; therefore, for Mn in MnTe ($S=5/2$), the Hamiltonian reduces to three terms:
\begin{equation}
	\mathcal{H} = B_2^0O_2^0 + B_4^0O_4^0 + B_4^3O_4^3 .
\end{equation}
Defining $\theta$ as the angle between the \neel\ vector and the crystallographic $c$-axis, and $\phi$ as the azimuthal angle from the $a$-axis, the Hamiltonian takes the form
\begin{multline}
    \mathcal{H} = B_2^0S^2\left[3\cos^2\theta-1\right] \\
    + B_4^0S^4\left[35\cos^4\theta-30\cos^2\theta+3\right] \\
    + B_4^3S^4\left[\sin^3\theta\cos\theta\cos\left(3\phi\right)\right] .
\end{multline}
Evidently, the $B_2^0$ and $B_4^0$ terms solely dictate whether the system exhibits easy-axis ($c$-axis) or easy-plane ($ab$-plane) anisotropy.
Because experiments confirm the magnetic moments lie predominantly in the basal plane, we substitute $\theta=\pi/2 + \delta$, where $\delta$ represents a small out-of-plane canting.
Expanding to leading order in $\delta$ yields
\begin{multline}\label{eq:Hamiltonian}
		\mathcal{H} = B_2^0S^2\left[3\delta^2-1\right] - B_4^0S^4\left[30\delta^2-3\right] \\
        - B_4^3S^4\left[\delta\cos\left(3\phi\right)\right] \\
        \propto K_\text{out}\delta^2 - K_\text{in}\delta\cos\left(3\phi\right),
\end{multline}
where $K_\text{out}$ and $K_\text{in}$ are the effective anisotropy constants governing the out-of-plane and in-plane alignments, respectively.

When $\delta=0$, the leading-order anisotropic contribution in Eq.~\ref{eq:Hamiltonian} vanishes.
Consequently, we must evaluate whether $\delta$ is strictly zero in this model.
Minimizing the energy with respect to $\delta$ yields the optimal canting angle
\begin{equation}
	\delta_\text{opt} = \frac{K_\text{in}\cos\left(3\phi\right)}{2K_\text{out}} .
\end{equation}
This is our first key result: the magnetic moments must cant toward the $c$-axis, with the magnitude of this tilt depending on both the in-plane azimuthal orientation and the ratio of the two anisotropy constants.
Substituting $\delta_\text{opt}$ back into Eq.~\ref{eq:Hamiltonian} reveals that the effective Hamiltonian varies as $\cos\left(6\phi\right)$.
This is our second key result: the single-ion anisotropy for spins near the basal plane exhibits a sixfold azimuthal symmetry.

We now relate the $B_n^m$ coefficients to the local crystal geometry.
In MnTe, the Mn atoms are octahedrally coordinated by six surrounding Te atoms; the critical geometric parameter is the angle $\alpha$ between the Mn--Te bond and the $c$-axis.
Based on the MnTe lattice parameters $a$ and $c$, this angle is given by
\begin{equation}
	\begin{split}
		\cos\alpha & = \frac{c}{4l} ,\\
		\sin\alpha & = \frac{a}{l\sqrt{3}} ,
	\end{split}
\end{equation}
where $l$ is the Mn--Te bond length,
\begin{equation}
	l=\sqrt{\frac{a^2}{3}+\frac{c^2}{16}} .
\end{equation}
Within a point-charge model, one can derive scaling relations for the $B_n^m$ coefficients.
For this analysis, we neglect the $B_4^0$ term by assuming the Mn $3d$ orbitals are well localized relative to the Mn--Te bond length.
Since these coefficients scale as $\langle r^n \rangle / l^{n+1}$ (where $\langle r \rangle$ is the average orbital radius), $B_4^0 \ll B_2^0$.
The leading coefficients therefore scale as~\cite{hutchings1964point}
\begin{equation}
	\begin{split}
		B_2^0 & \propto \frac{3\cos^2\alpha-1}{l^3} = \frac{1}{l^5}\left(\frac{c^2}{8}-\frac{a^2}{3}\right) ,\\
		B_4^3 & \propto \frac{\sin^3\alpha\cos\alpha}{l^5} = \frac{1}{l^9}\frac{a^3c}{12\sqrt{3}} .
	\end{split}
\end{equation}

The term in the Hamiltonian governing the in-plane azimuthal dependence is
\begin{multline}
		K_\text{in}\delta\cos\left(3\phi\right) = \frac{K_\text{in}^2\left[1+\cos\left(6\phi\right)\right]}{4K_\text{out}} \\
        \propto \frac{\left(B_4^3\right)^2\cos\left(6\phi\right)}{B_2^0} ,
\end{multline}
where we have neglected $B_4^0$ and omitted constant prefactors.
Substituting our scaling relations for $B_n^m$ yields
\begin{equation}
	\frac{\left(B_4^3\right)^2}{B_2^0} \propto \frac{a^6c^2}{l^{13}\left(3c^2-8a^2\right)} .
\end{equation}
Because $a$, $c$, and $l$ are strictly positive, the sign of this expression flips precisely when the lattice-parameter ratio crosses the critical value $c/a=\sqrt{8/3}\simeq1.633$.
Since this prefactor scales the $\cos\left(6\phi\right)$ energy landscape, this sign reversal triggers a $30^\circ$ rotation of the \neel\ vector within the basal plane.
This constitutes our final key result: the temperature-dependent evolution of the lattice parameters dictates a critical point at which the magnetic easy axis shifts from parallel to the $a$-axis to $30^\circ$ away from it, in direct agreement with our experimental observations.

Reintroducing the $B_4^0$ term demonstrates that its inclusion merely shifts the precise $c/a$ ratio at which the \neel\ vector reorients, without fundamentally altering the physical mechanism.
Although the exact relative strengths of the $B_2^0$ and $B_4^0$ terms remain unknown, the effective anisotropy ratio strictly obeys the form
\begin{equation}
	\frac{K_\text{in}^2}{K_\text{out}} \propto f\left(c^2\right) - g\left(a^2\right) ,
\end{equation}
where the proportionality constant maintains a fixed sign.
Consequently, while this model cannot predict the exact critical $c/a$ ratio \textit{a priori} without experimental constraints on the individual term strengths, the overarching conclusion remains highly robust.

\subsection*{Dipolar anisotropy}
Another potential source of magnetic anisotropy is the dipole-dipole interaction between Mn moments.
Numerical calculations of these energies confirm that while this interaction strongly favors in-plane alignment, lowering the energy by approximately 100~$\mu$eV/$\mu_\text{B}^2$ per Mn relative to the $c$-axis orientation, it introduces zero anisotropy within the basal plane.

\end{document}